\documentclass[12pt]{iopart}
\usepackage[dvips]{color}
\usepackage{epsf}
\usepackage{ulem}

\begin{document}

\title{Foreign exchange market: return distributions, multifractality, anomalous multifractality and Epps effect}

\author{Stanis{\l}aw Dro\.zd\.z$^{1,2}$, Jaros{\l}aw Kwapie\'n$^{1}$, Pawe{\l} O\'swi\c ecimka$^{1}$ and Rafa{\l} Rak$^{2}$}

\address{$^1$ Institute of Nuclear Physics, Polish Academy of Sciences, ul. Radzikowskiego 152, PL - 31-342 Krak\'ow, Poland\\
$^2$ Faculty of Mathematics and Natural Sciences, University of Rzesz\'ow, PL - 35-310 Rzesz\'ow, Poland }

\ead{Stanislaw.Drozdz@ifj.edu.pl}

\begin{abstract}

We present a systematic study of various statistical characteristics of high-frequency returns from the foreign exchange market. This study is based on six exchange rates forming two triangles: EUR-GBP-USD and GBP-CHF-JPY. It is shown that the exchange rate return fluctuations for all the pairs considered are well described by the nonextensive statistics in terms of $q$-Gaussians. There exist some small quantitative variations in the nonextensivity $q$-parameter values for different exchange rates (which depend also on time scales studied) and this can be related to the importance of a given exchange rate in the world's currency trade. Temporal correlations organize the series of returns such that they develop the multifractal characteristics for all the exchange rates with a varying degree of symmetry of the singularity spectrum $f(\alpha)$ however. The most symmetric spectrum is identified for the GBP/USD. We also form time series of triangular residual returns and find that the distributions of their fluctuations develop disproportionately heavier tails as compared to small fluctuations which excludes description in terms of $q$-Gaussians. The multifractal characteristics for these residual returns reveal such anomalous properties like negative singularity exponents and even negative singularity spectra. Such anomalous multifractal measures have so far been considered in the literature in connection with the diffusion limited aggregation and with turbulence. Studying cross-correlations among different exchange rates we find that market inefficiency on  short time scales leads to the occurrence of the Epps effect on much longer time scales, though comparable to the ones for the stock market. Although the currency market is much more liquid than the stock markets and it has much larger transaction frequency, the building-up of correlations takes up to several hours - time that does not differ much from what is observed in the stock markets. This may suggest that non-synchronicity of transactions is not the unique source of the observed effect.

\end{abstract}

\maketitle

\section{Introduction}

The Foreign Exchange market (FX), with its daily volume of over five trillions USD in 2009 is by far the world's largest financial market. Any other financial market can hardly approach such volume. This market connects international institutions participating in currency exchange transactions all across the world and encompasses essentially everything of what is going on in the world, first of all including economic factors, political conditions and market psychology, all of them constantly changing. Also, this market has direct influence on all other markets because any price is expressed in terms of a currency. The large volume makes it virtually impossible to control from outside and there is no friction (transactions are basically commission free). Due to time differences FX transactions are performed 24 h a day, 5 days a week with maximum volume between 13:00 and 16:00 GMT, when both American and European markets are open. Hence, the FX time series relations represent an exceptionally complex network indeed, and they therefore constitute an especially challenging target of a detailed quantitative analysis.

In connection with the almost continuous trading, FX is also much more effective and liquid than other speculative markets. A significance of this market (an example of globalization) is even more important, since it became an indicator of a condition of the world's economy. From a physicist's viewpoint FX is a complex system with extremely convoluted time dependencies. The FX effectiveness is intensified by the correlations between exchange rates known as the triangular arbitrage. It is possible only on small time scales and disappears immediately after taking advantage of inconsistent crossrates by the traders. To quantify such correlations we employed the multifractal analysis measuring nonlinear features of time series, in particular their multifractal spectra. Especially interesting is a relation between the fractal properties of the exchange rates remaining in the triangular dependency. Encapsulating this relation - especially of the empirical residuals within the triangle - may shed more light on this so far poorly understood issue. The resulting scale free statistics encodes information about complex interactions in FX.

The FX data~\cite{data} used in the present analysis include the following six indicative exchange rate pairs: USD/EUR, EUR/GBP, GBP/USD, JPY/GBP, GBP/CHF, and CHF/JPY sampled with 1 minute frequency over the period from 21:00 on January 2, 2004, until 21:00 on March 30, 2008 (1183 days, 169 weeks). The selection of currency pairs and time interval was constrained by availability of the sufficient quality data, so not all important pairs could be included in our analysis. In a consequence, we deal with six synchronized time series of length $T=1,703,520$ that can be labeled as $x_A^B(t_i)$, $i=1,...,T$, where $x(t_i)$, denotes a value of currency $A$ expressed in terms of a currency $B$ at time $t_i$. Consequently, the corresponding returns over the time period $\Delta t$ are expressed as
\begin{equation}
G_A^B(t_i; \Delta t)= \ln x_A^B(t_i+\Delta t) - \ln x_A^B(t_i).
\label{G}
\end{equation}
Let us define residual returns as
\begin{equation}
G^{\bigtriangleup}(t_i;\Delta t) = G_A^B(t_i;\Delta t) + G_B^C(t_i;\Delta t) + G_C^A(t_i;\Delta t),
\label{residual}
\end{equation}
which are expected to fulfill the following relation:
\begin{equation}
G^{\bigtriangleup}(t_i;\Delta t) = 0.
\label{triangle}
\end{equation}
Departures from (\ref{triangle}) generate the so-called triangular arbitrage opportunities that, whenever detected, may be exploited and in fact are commonly used for the risk free profit generation. In the contemporary markets an execution of the corresponding operation typically takes at most a few seconds~\cite{Aiba2003} and is thus far below the scales of 1 min. considered here. Of course, this does not yet imply that returns of the corresponding three exchange rates synchronously evaluated at larger time scales $(\Delta t)$ obey Eq.~(\ref{triangle}) exactly. Viewed at the same instant of time some mismatch may result just from the time needed (a few seconds) to execute the arbitrage opportunity and this introduces some dispersion. It may also reflect some noise component involved. Clearly, on the larger time scales such effects of departure from zero in Eq.~(\ref{triangle}) become less and less relevant relative to the total return. For the two exchange rate triangles, EUR-GBP-USD and GBP-CHF-JPY, that operate within the six currency exchange rates considered here, the logarithmic returns defined by Eq.~(\ref{G}) are shown in Fig.~1 for all the three exchange rates involved in each triangle accompanied by the sum of these three returns (the left hand side of Eq.~(\ref{triangle})). These characteristics are presented for the shortest time lag here accessible of $\Delta t = 1$ min (left panels) and for $\Delta t = 60$ min (right panels). As far as the magnitude of the fluctuations is concerned one sees essentially no sizeable difference for the four series so generated for $\Delta t = 1$ min. The only visibly detectable difference is in the structure of fluctuations; the sum of the three returns (lowest row in each panel) looks more uniform in each case and large fluctuations are less frequent. The situation changes considerably for $\Delta t = 60$ min. What is natural, the magnitude of returns for the individual exchange rates significantly increases while, at the same time, their sum in each triangle decreases even relative to $\Delta t = 1$ min. What seems also worth pointing already at this stage is that the fluctuations of returns in the EUR/GBP exchange rate look more quiet than for the remaining two pairs in this triangle for both time lags of 1 and 60 min. The same applies to the CHF/GBP exchange rate in the second triangle.

\begin{figure}
\hspace{-0.5cm}
\epsfxsize 10cm
\epsffile{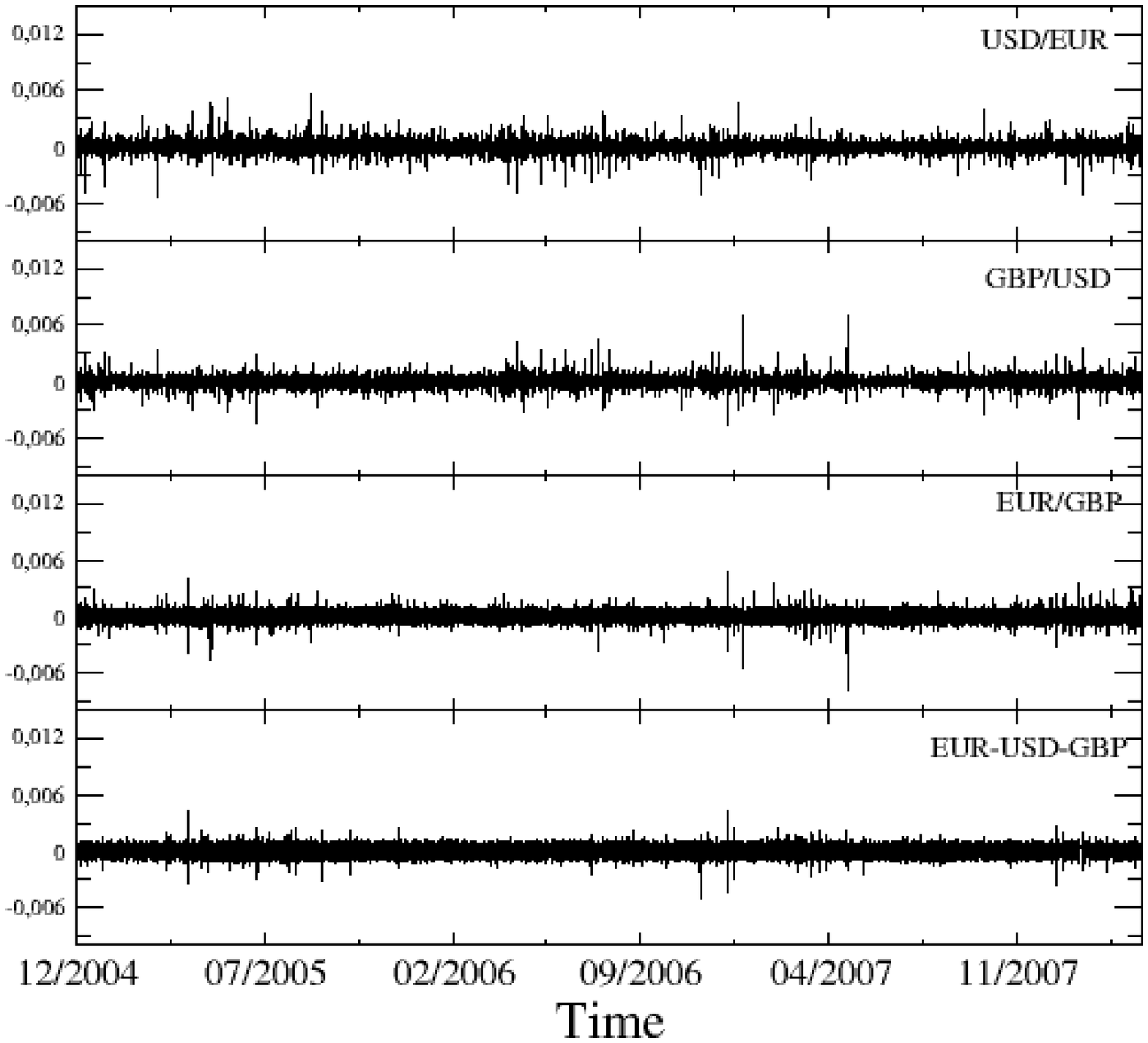}
\hspace{-2.0cm}
\epsfxsize 10cm
\epsffile{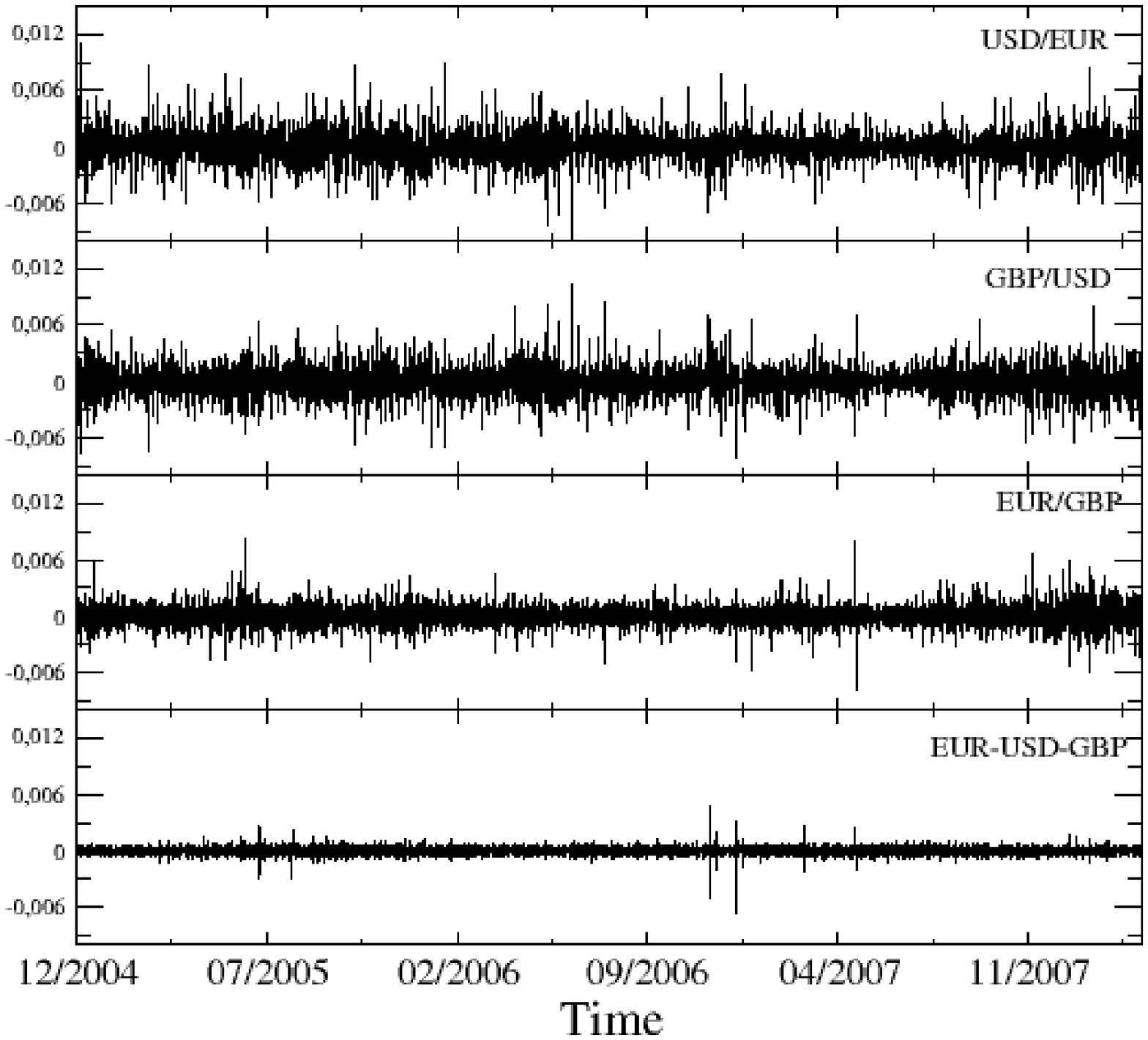}

\vspace{-0.5cm}
\hspace{-0.5cm}
\epsfxsize 10cm
\epsffile{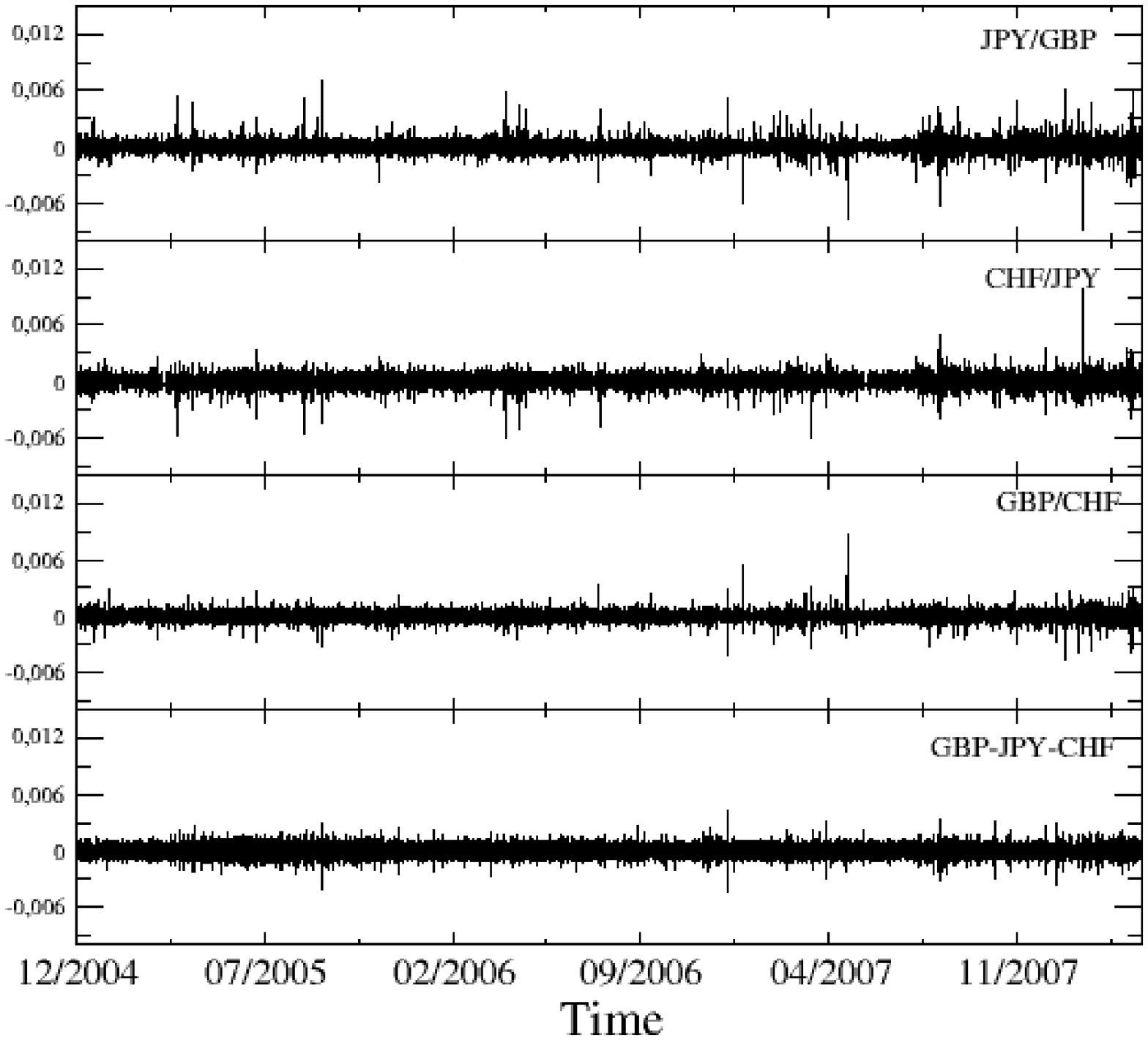}
\hspace{-2.0cm}
\epsfxsize 10cm
\epsffile{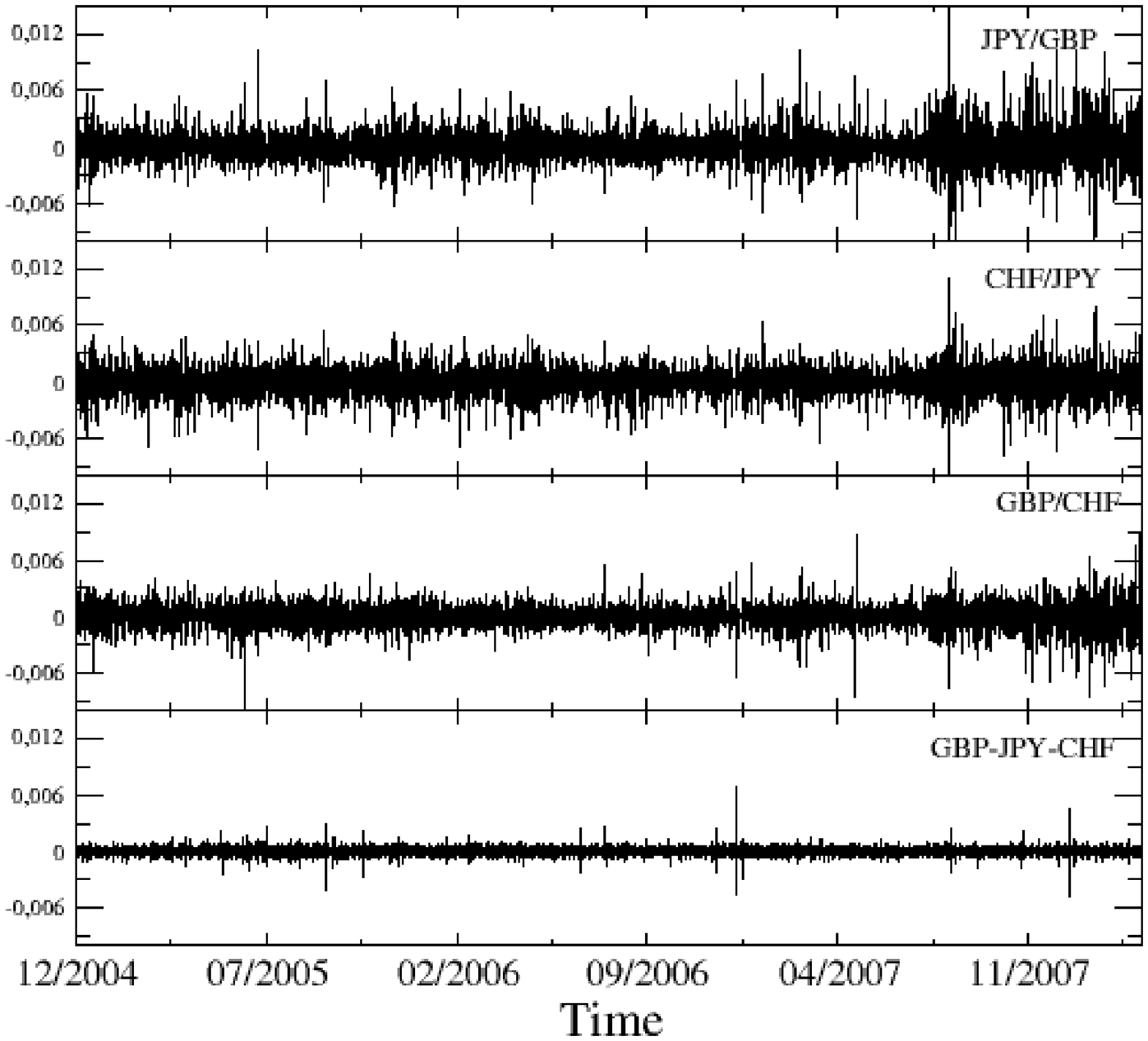}
\caption{Time series of 1 min. (left) and 60 min. (right) returns for six currency exchange rates forming two triangles: EUR-GBP-USD (top) and GBP-CHF-JPY (bottom). In each case the appropriate residual time series (\ref{residual}) is also shown in the lowermost panel.}
\label{fig.1}
\end{figure}

Irrespective of such details, the relation expressed by Eq.~(\ref{triangle}) definitely introduces a crucial factor that affects the dynamics of the currency exchange network in multiple ways. It first of all sets constraints on the dynamics by effectively reducing the number of independent degrees of freedom. For $N$ currencies instead of $N(N-1)/2$ there in fact exists $N-1$ independent exchange rates. This crucially shapes topology of the corresponding exchange rate network structure~\cite{Gorski2008,Kwapien2009a,Kwapien2009b}. Furthermore, some of the exchange rates may be primarily driven by the trade needs or some speculation specific arguments while dynamics of the others may be affected more by the market adjustments towards eliminating or at least reducing the arbitrage opportunities.

In the following we analyze a few most informative statistical characteristics of time series for the exchange rates listed above. These characteristics determine the sectional organization of the paper. In each section some novel results not discussed previously in literature are presented. In Section 2 we show that although the distribution of returns for the individual exchange rates can be approximated (similar to the returns from other financial markets) by the $q$-Gaussian distributions, the residual signals $G^{\bigtriangleup}(t_i;\Delta t)$ clearly do not. Section 3 is devoted to an analysis of temporal correlations and detection of repeatable patterns of market activity. We found that exchange rates exhibit different temporal correlation properties depending on the trading significance of a particular currency pair. Next, in Section 4 we study multifractal properties of the exchange rates in terms of the singularity spectra and, for the first time, identify their anomalous structure: negative exponents $\alpha$ and negative $f(\alpha)$. Finally, Section 5 deals with the cross-correlation structure of the currency triangles, documenting occurrence of the Epps effect, i.e. increase of coupling strength between the exchange rates with increasing sampling time $\Delta t$, observed for surprisingly long $\Delta t$.

\section{Distribution of return fluctuations}

One of the most relevant quantitative characteristics of the financial dynamics is the functional form of the distribution of returns. The related - in the past well identified stylized fact - is the so-called inverse cubic power-law~\cite{Gabaix2003} which applies to developed stock markets~\cite{Lux1996,Gopi1999,Plerou1999,Drozdz2003}, to some emerging stock markets like the Polish market~\cite{Rak2007}, to the commodity market~\cite{Matia2002}, as well as to the most traded currency exchange rates~\cite{Muller1995} in the early 1990s. Of course, this type of distribution is L\'evy-unstable and thus for the sufficiently long time lags $\Delta t$ the returns distribution is expected to converge towards a Gaussian. This convergence, and thus departures from the inverse cubic power-law, has been found to be very slow as a function of $\Delta t$. In the more recent data, however, this convergence appears~\cite{Drozdz2003,Rak2007} significantly faster and departures from the inverse cubic power-law in the contemporary stock markets can be seen already for $\Delta t = 1$ min.

A formalism that appears~\cite{Rak2007,Drozdz2009} attractively compact and economic for describing the two extremes - the inverse cubic as well as the Gaussian distributions - including all the intermediate cases is the one based on the generalized nonextensive entropy~\cite{Tsallis1988}. Accordingly, optimization of the corresponding generalized entropic form under appropriate constraints~\cite{Tsallis1995} yields the following $q$-Gaussian form for the distribution of probabilities
\begin{equation}
p\left( x\right) =\mathcal{N}_{q}\,e_{q}^{-\mathcal{B}_{q}\left( x-\bar{\mu}_{q}\right) ^{2}},
\label{px}
\end{equation}
where the constants $\mathcal{N}_{q}$, $\mathcal{B}_{q}$, $\bar{\mu}_{q}$ and the $q$-exponential function $e_{q}^x$ are defined in Appendix. In order to attain a better stability in this kind of the analysis, we prefer to use cumulative form of the distribution (\ref{px}) - see Appendix for the corresponding formulae.

As a standard procedure that makes the distributions for different exchange rates directly comparable, we convert $G_A^B$ of Eq.~(\ref{G}) into the normalized returns $g_A^B$ defined as
\begin{equation}
g_A^B = {G_A^B - \langle G_A^B \rangle_T \over v_A^B},
\label{g}
\end{equation}
where $v_A^B \equiv v_A^B(\Delta t)$ is the standard deviation of returns over the period $T$.

\begin{figure}
\hspace{0.0cm}
\epsfxsize 14cm
\epsffile{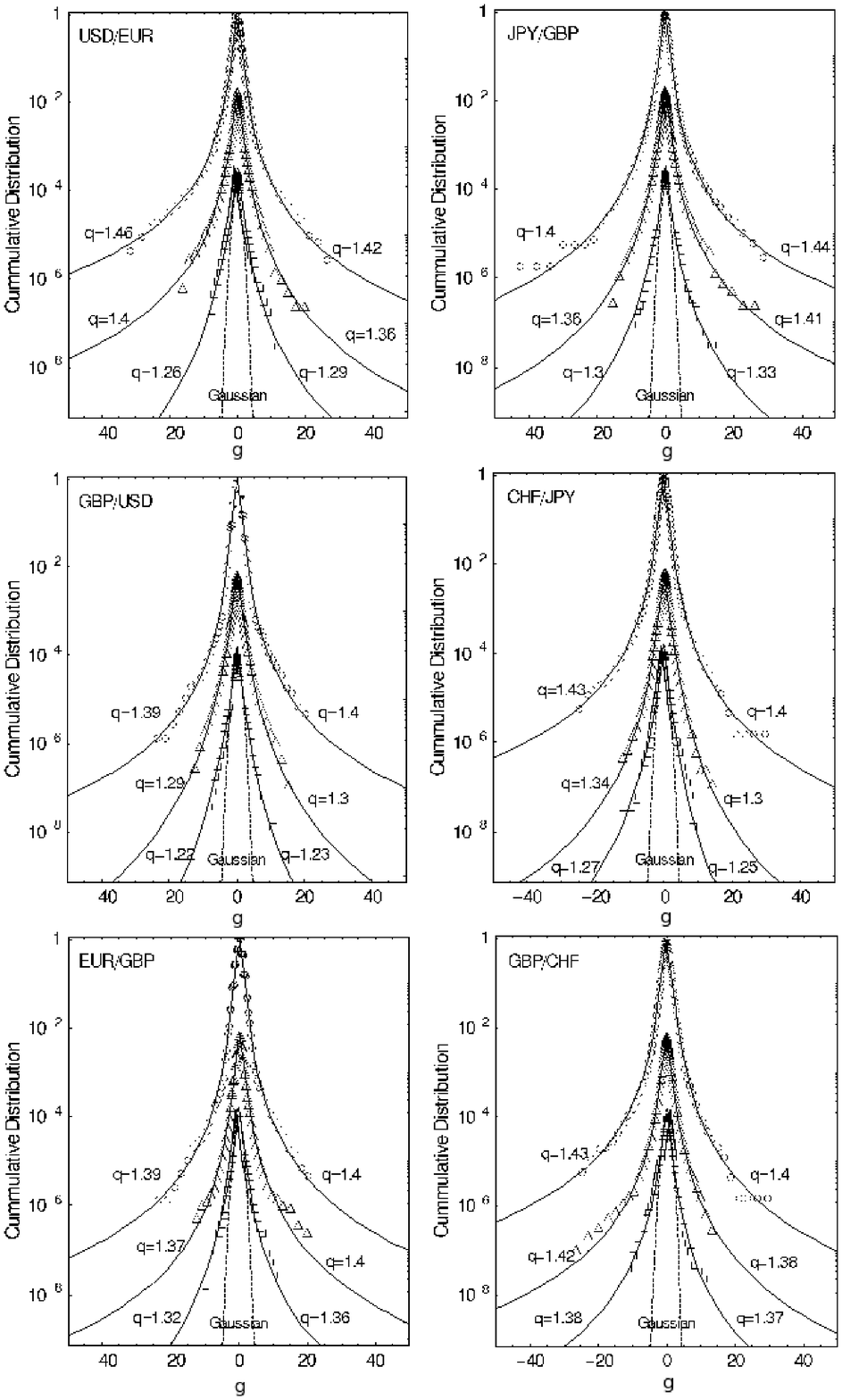}
\caption{Cumulative distributions of 1 min., 10 min. and 60 min. returns for all the exchange rates from the triangles EUR-GBP-USD (left) and GBP-CHF-JPY (right). In each case the empirical distributions are best-fitted by the $q$-Gaussian distributions.}
\label{fig.2}
\end{figure}

The empirical cumulative distributions for all the exchange rates considered here versus their best fits in terms of the $q$-Gaussians (Eq.~(\ref{Pcx})) are shown in Fig.~2. The left column corresponds to the three exchange rates form the EUR-GBP-USD triangle while the right column to the exchange rates from the GBP-CHF-JPY triangle. As one can see in all the cases the $q$-Gaussians provide a very reasonable representation over the whole span of fluctuations and for the increasing return time-lags $\Delta t$ of 1 min., 10 min. and 60 min. Some asymmetry between the left and right wings in the distributions, as expressed by the slightly different values of the corresponding $q$-parameter can be detected. As expected, with the increasing time-lags $\Delta t$ the $q$-values decrease which reflects an expected (slow) convergence to the Gaussian ($q=1$) distribution. Among the pairs considered here one interesting difference in this respect can be detected however: the decrease of the $q$ values with increasing $\Delta t$ is slower for the intra-European exchange rates (EUR/GBP and GBP/CHF) than for the intercontinental ones. The convergence towards a Gaussian distribution is thus slower in the former case. This result is qualitatively similar to an earlier observation based on data from 1992-93 (i.e. long before the introduction of euro), which was documented in ref.~\cite{gillaume97}. This indicates that the global forex market is largely stable as regards the statistical distribution of returns. Moreover, the degree of convergence towards the Gaussian distribution appears to behave similarly as in the contemporary stock markets~\cite{Drozdz2003,Drozdz2009}. It also needs to be noticed that analogously as the recent S\&P500 analysis shows~\cite{Drozdz2007} a slight departure from the inverse cubic power law (which in the present formalism of the $q$-Gaussians corresponds to $q=3/2$) takes place already for $\Delta t = 1$ min. Perhaps this law is approached more accurately only for the time-lags even smaller than 1 min.

\begin{figure}
\hspace{0.0cm}
\epsfxsize 17cm
\epsffile{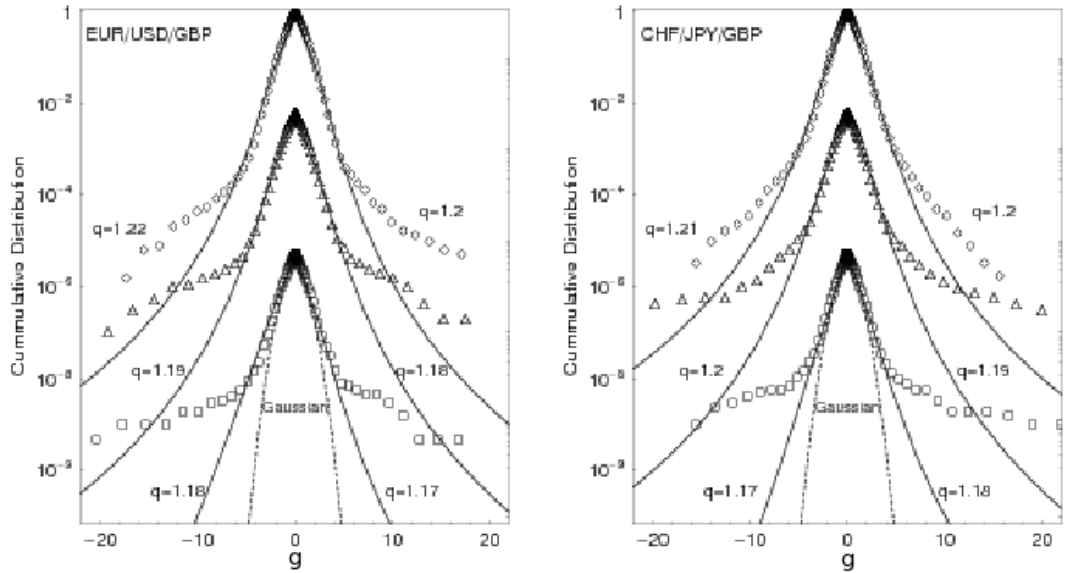}
\caption{Cumulative distributions of normalized residual returns $g^{\bigtriangleup}(t;\Delta t)$ for $\Delta t=1$ min., $\Delta t=10$ min., and $\Delta t=60$ min. corresponding to the EUR-GBP-USD (left) and GBP-CHF-JPY (right) triangles. Central parts of empirical distributions are best-fitted by $q$-Gaussians.}
\label{fig.3}
\end{figure}

It is interesting to look at the distributions of the residual returns as defined by Eq.~(\ref{triangle}), since - according to our knowledge - they have not yet been shown in literature. Fig.~3 shows such distributions for the same time scales: $\Delta t = 1$, 10 and 60 min. The $q$-Gaussians are now able to fit only the central part of the distributions up to about three mean standard deviations. The corresponding $q$-values only weakly decrease with $\Delta t$ starting from $q \approx 1.2$ for $\Delta t = 1$ min. The tails of the empirical distributions are disparately thicker than expected by the $q$-Gaussian model and become even more such for larger $\Delta t$. Such an effect is in fact visible already in Fig.~1. The background fluctuations in their lowest panels sizeably decrease with increasing $\Delta t$ while at the same time the largest fluctuations remain of the same order as compared to the individual exchange rates shown in the upper panels. These 'outliers' may reflect a longer time needed to balance departures from Eq.~(\ref{triangle}) resulting from sudden large returns in one of the pairs in the triangle. These characteristics are very similar in both the triangles considered here.

\section{Temporal correlations}

The issue of the character of temporal correlations is equally important in the financial context and many related questions still remain open. The simplest measure of the temporal correlations is in terms of the autocorrelation function $c(\tau)$ which for a function $f(t)$ is defined as
\begin{equation}
c(\tau) = \langle f(t + \tau) f(t) \rangle,
\label{c}
\end{equation}
where $\langle ... \rangle$ denotes an average over $t$. The most studied cases in the financial context correspond to the autocorrelation of returns, here represented by $g^A_B$, and of the volatility which can be defined as the modulus $\vert g^A_B \vert$ of returns.

\begin{figure}
\hspace{2.0cm}
\epsfxsize 12cm
\epsffile{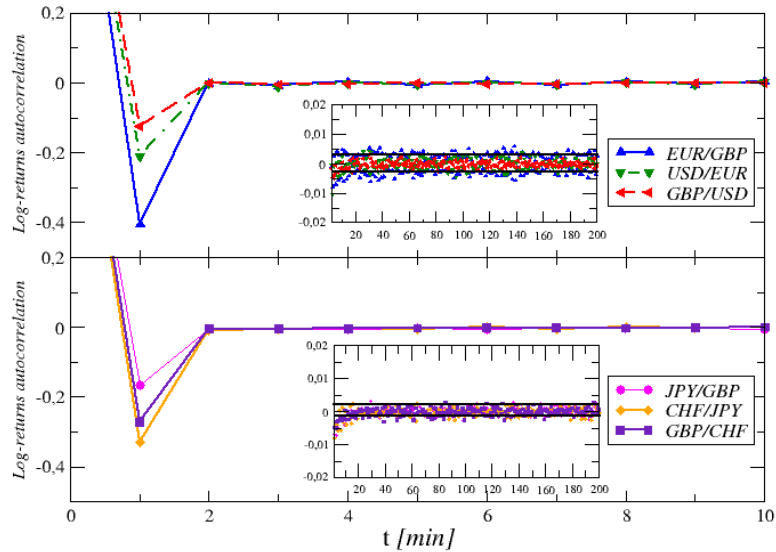}
\caption{Autocorrelation function (\ref{c}) of 1 min. returns corresponding to each exchange rate from the two considered currency triangles. Insets: the same function for a broader range of $\tau$ and with rescaled vertical axis; the 95\% confidence levels are also indicated.}
\label{fig.4}
\end{figure}

Fig.~4 shows the return autocorrelation as a function of $\tau$, in the upper panel for the exchange rates from the EUR-GBP-USD triangle and in the lower panel for the exchange rates from the GBP-CHF-JPY triangle. Similarly as for the typical stock market returns, in fact even faster because in all these six cases already for $\tau = 2$ min., such an autocorrelation is seen to assume values at the null level. This even faster disappearance of the FX return autocorrelation is probably related to a higher liquidity involved. Characteristic for all the cases considered here is appearance of the 'correlation hole'~\cite{Drozdz1995} of negative autocorrelation for $\tau = 1$ min. which reflects some anti-persistent tendencies on such short time scales. There is no unique explanation of this effect first mentioned in ref.~\cite{goodhart91}; it can originate from the divergent opinions of traders about the direction of imminent price changes as well as from certain actions of market-makers and banks~\cite{bollerslev93,bollerslev94}. The depths of these 'holes' is different for different pairs even within the same triangle. For the pairs that can be considered leading in the FX dynamics (GBP/USD, USD/EUR, and JPY/GBP) this depth at $\tau = 1$ min. can be seen to be smaller than for the other (EUR/GBP, GBP/CHF, CHF/JPY) pairs.

The volatility autocorrelations for the same six pairs of currencies are shown in Fig.~5. It is quite obvious, due to the log-log scale used in this Figure, that their behavior can well be approximated by the power law time dependance $c(\tau) \sim \tau^{\alpha}$ with $\alpha \approx 0.4$. This value of the scaling index does not differ from the value which is typical for a majority of stock markets~\cite{Gopi1999,Ding1993,Bouchaud2000}. Even more, all the volatility autocorrelations between events that are separated by more than about $10^4$ basic units (1 min.) suddenly drop down and start oscillating between the positive and negative values (not visible in this Figure) with a decreasing amplitude. This effect has recently been found also for stock markets in~\cite{Drozdz2009}. Its natural interpretation is that the so-determined time horizon of the power-law volatility autocorrelations corresponds to an average length of either low or high volatility clusters. As far as the FX exclusively characteristics are concerned, one more effect needs to be pointed out based on Fig.~5. For the same pairs (GBP/USD, USD/EUR, JPY/GBP) that above have been indicated as the leading ones in the FX dynamics, the volatility autocorrelation is systematically stronger than for the remaining pairs (as the relative location of the corresponding lines shows).

\begin{figure}
\hspace{2.0cm}
\epsfxsize 12cm
\epsffile{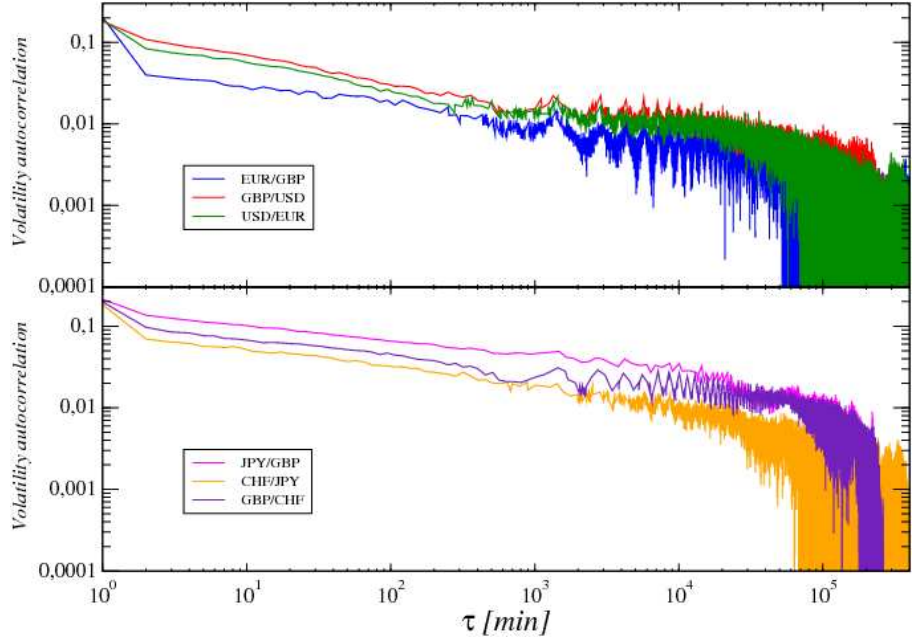}
\caption{Autocorrelation function (\ref{c}) of volatility time series corresponding to each exchange rate from the two considered currency triangles. Daily trend has been removed according to a standard procedure, in which at each instant signal is divided by the volatility mean standard deviation characteristic for this particular instant (as evaluated from all the trading days included).}
\label{fig.5}
\end{figure}

A somewhat more advanced method to quantify the character of financial temporal correlations is to use a variant of the correlation matrix. In this approach, initiated in ref.~\cite{Kwapien2000,Kwapien2002}, the entries of the corresponding matrix are the correlation coefficients between the time series of returns representing different disconnected time-intervals, like the consecutive trading days or weeks. The structure of eigenvalues and eigenvectors of such a matrix allows then to quantify further characteristics of the temporal correlations.

Suppose from the time series $g(t_i)$ of length $T$ one extracts $K$ disconnected series $g_{\beta}(t_i)$ $(\beta =1,...,K)$ of length $T_K$. Of course, the condition $K T_K \le T$ has to be fulfilled. By using such time series as rows one forms an $K \times T_K$ matrix $\bf M$. Then, the correlation matrix is defined as $ {\bf C} = (1/ T) \ {\bf M} \tilde{{\bf M}}$, where $\tilde{\cdot}$ is matrix transpose. By diagonalizing $\bf C$:
\begin{equation}
{\bf C} {\bf v}^k = \lambda_k {\bf v}^k, \label{C}
\end{equation}
one obtains the eigenvalues $\lambda_k$ $(k=1,...,K)$ and the corresponding eigenvectors ${\bf v}^k = \{ v_{\beta}^k \}$. In the limiting case of entirely random correlations, the density of eigenvalues $\rho_C(\lambda)$ is known explicitely~\cite{Marcenko1967,Sengupta1999}, and reads:
\begin{equation}
\rho_C(\lambda) = {Q \over {2 \pi \sigma^2}} {\sqrt{ (\lambda_{max} - \lambda)
(\lambda -
\lambda_{min})} \over {\lambda}},
\label{rho}
\end{equation}
where
\begin{equation}
\lambda^{max}_{min} = \sigma^2 (1 + 1/Q \pm 2 \sqrt{1/Q}),
\label{lambdaminmax}
\end{equation}
with $\lambda_{\rm min} \le \lambda \le \lambda_{\rm max}$, $Q=T_K/K \ge 1$, and where $\sigma^2$ is equal to the variance of the time series.

For a better visualization, each eigenvector can be associated with the corresponding time series by the following expression:
\begin{equation}
z_k(t_i) = \sum_{\beta=1}^K v_{\beta}^{k} g_{\beta}(t_i),~~~~~~k = 1,...,K; ~~~ i=1,...,T_K.
\label{eigensignal}
\end{equation}
Thus, these new time series form orthogonal components into which the original signal $g_{\beta}(t_i)$ is decomposed. They reflect distinct patterns of oscillations common to all the time intervals labeled with $\beta$. These time series can therefore be called the eigensignals.

\begin{figure}
\hspace{0.0cm}
\epsfxsize 7cm
\epsffile{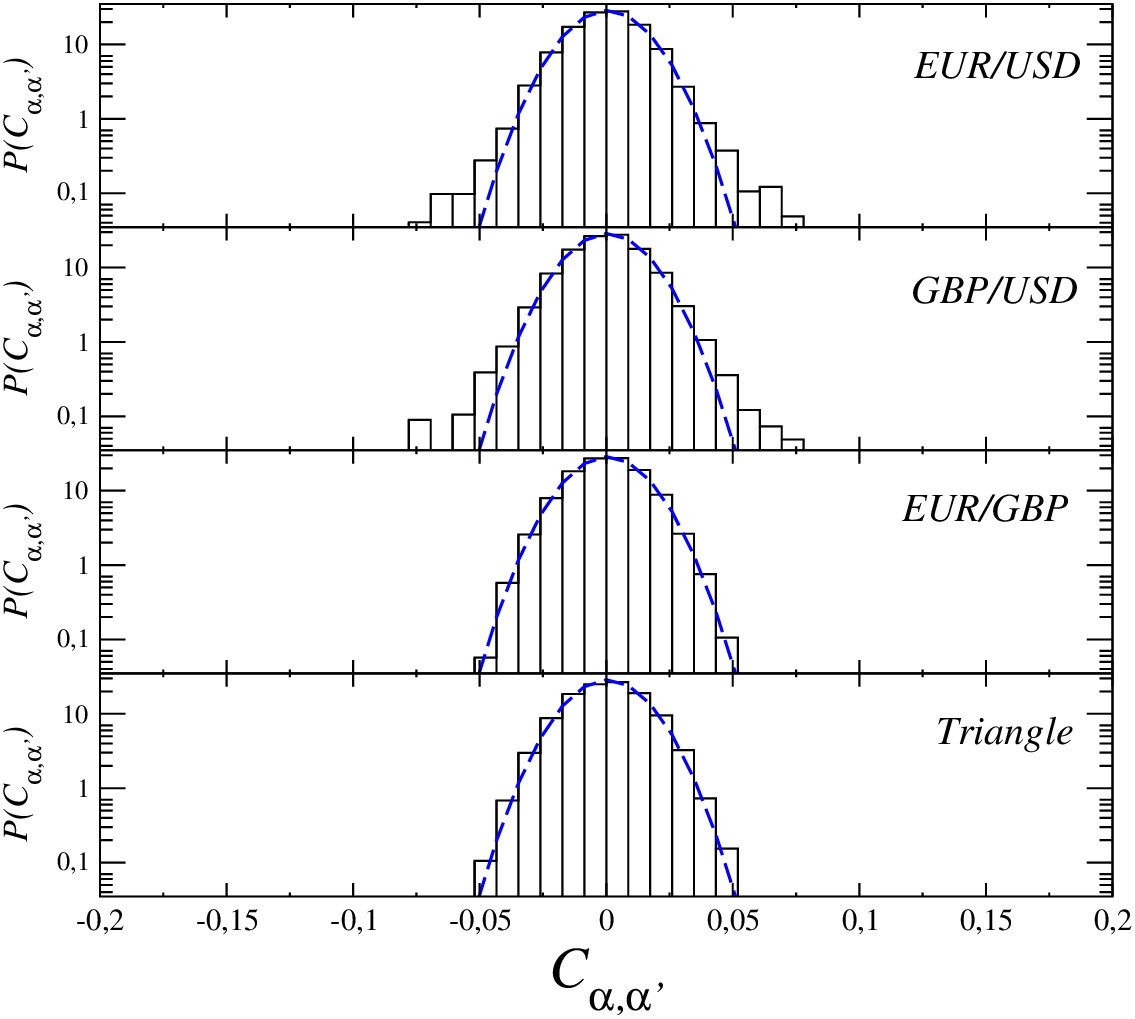}
\hspace{1.0cm}
\epsfxsize 7cm
\epsffile{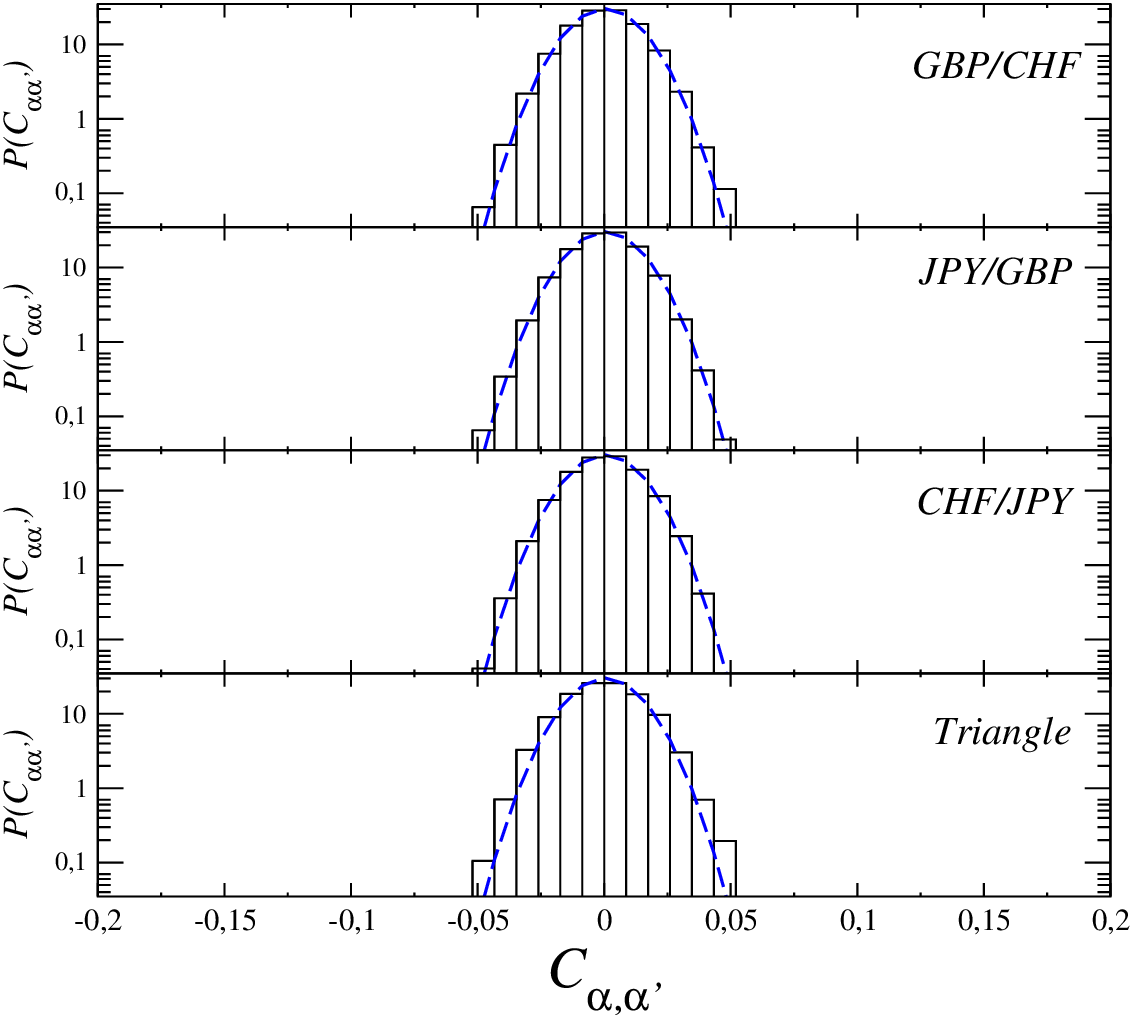}
\caption{Distributions of correlation matrix elements (histograms) together with the best-fitted Gaussian distributions (dashed lines).}
\label{fig.6}
\end{figure}

The above methodology on the weekly basis is now applied to the present FX data. Our original time series of returns comprise $K=169$ complete weeks counted from Sunday 21:00 until Friday 22:00 GMT. The length is $T_K=7,260$ min. For each pair of currencies and for the residual signals, the distributions of matrix elements are displayed in Fig.~6. As it can be seen, for a majority of the rates the empirical distributions are Gaussian-like. Only for the most heavily-traded pairs: USD/EUR and GBP/USD ${\bf C}$ has a significant number of non-Gaussian entries.

\begin{figure}
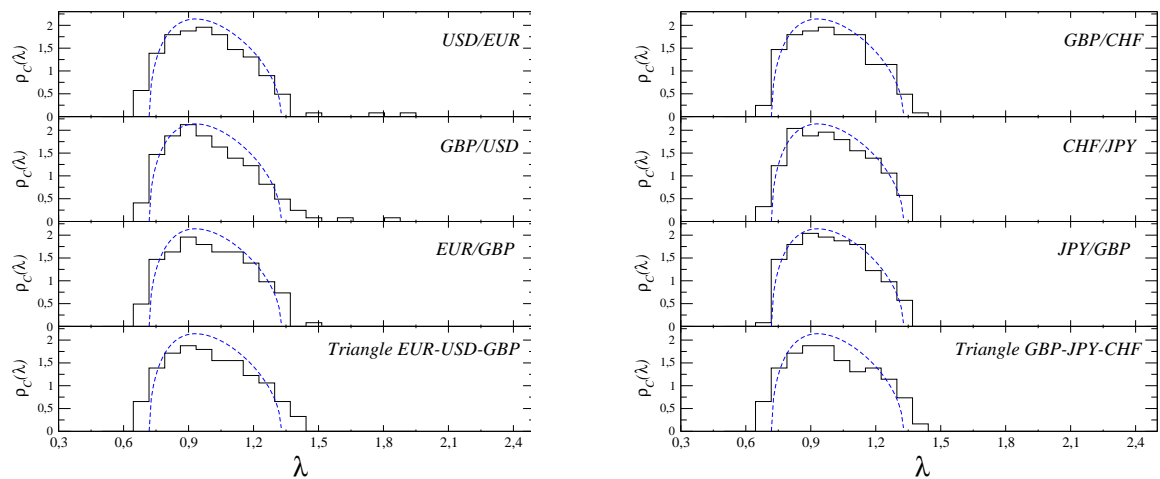

\hspace{0.0cm}
\epsfxsize 7cm
\epsffile{fig7a.eps}
\hspace{1.0cm}
\epsfxsize 7cm
\epsffile{fig7b.eps}
\caption{Density of correlation matrix eigenvalues corresponding to all the considered exchange rates and to the residual time series of triangle rule departures (bottom panels). Each empirical distribution (histograms) is associated with the theoretical distribution (dashed lines) described by Eq.~(\ref{rho}).}
\label{fig.7}
\end{figure}

Eigenvalue density for the corresponding correlation matrices for all the six currency exchange rates discussed here, including the residual time series representing departures from the triangle rule, are shown (histograms) in Fig.~7. The left panel corresponds to the EUR-GBP-USD triangle and the right panel to the GBP-CHF-JPY one. For comparison, the pure noise distribution - as prescribed (Eq.~(\ref{rho})) by the corresponding Wishart ensemble of random matrices~\cite{Marcenko1967,Sengupta1999} - is indicated by dashed lines. As one can see, besides some small departures at the edges, the empirical eigenvalue distributions do not differ much from their pure noise counterparts. This fact may indicate that the intra-week exchange rate behaviour does not involve any particularly significant repeatable patterns. This observation applies more to the pairs within the GBP-CHF-JPY triangle. The most evident departures between the empirical and the theoretical distributions one observes for the USD/EUR and GBP/USD exchange rates. In both these cases, the two largest eigenvalues stay visibly outside the noise range and thus may carry some system-specific information. This is in agreement with the distributions of matrix entries shown in Fig.~6.

\begin{figure}
\hspace{0.0cm}
\epsfxsize 9cm
\epsffile{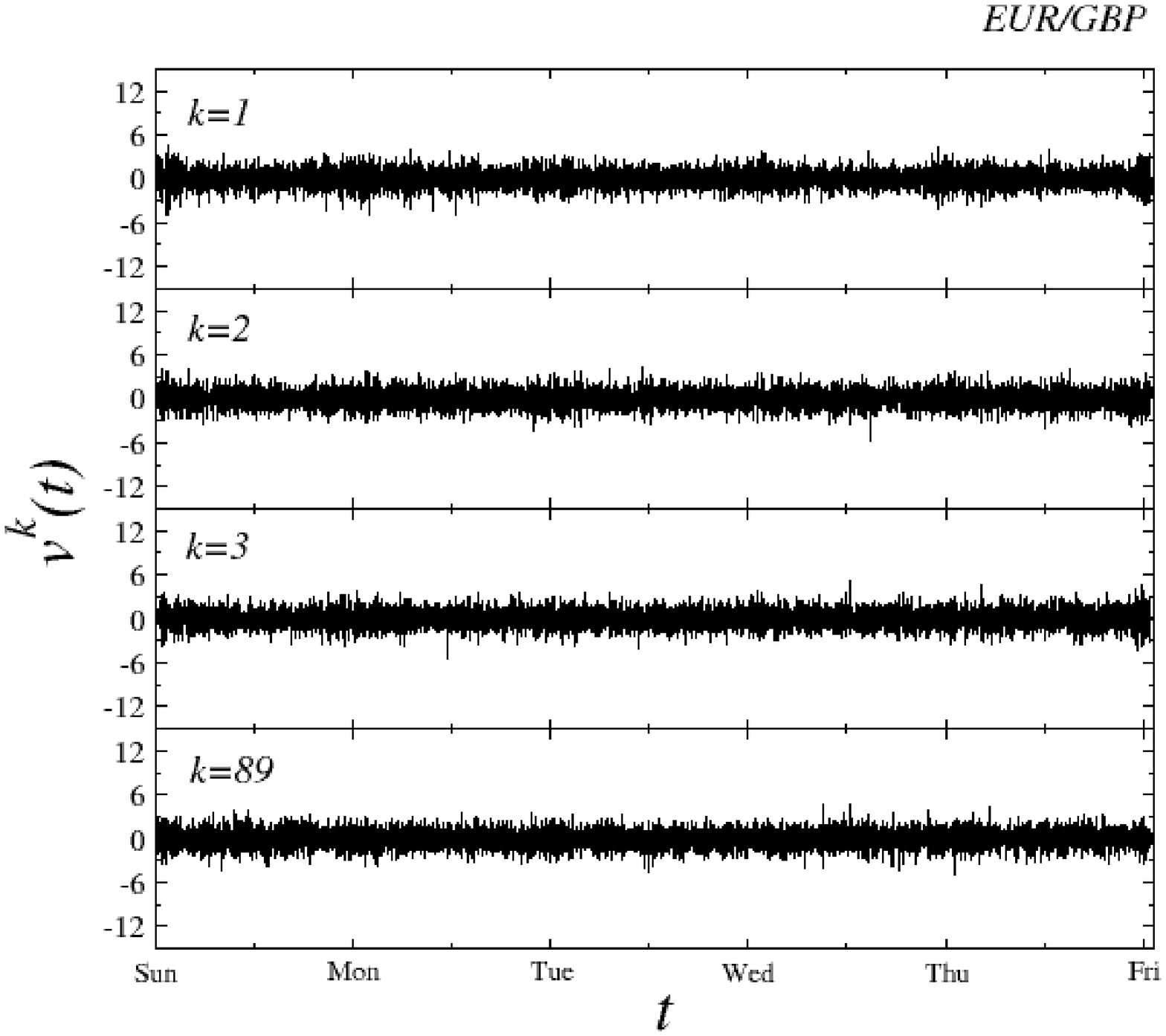}
\hspace{-1.5cm}
\epsfxsize 9cm
\epsffile{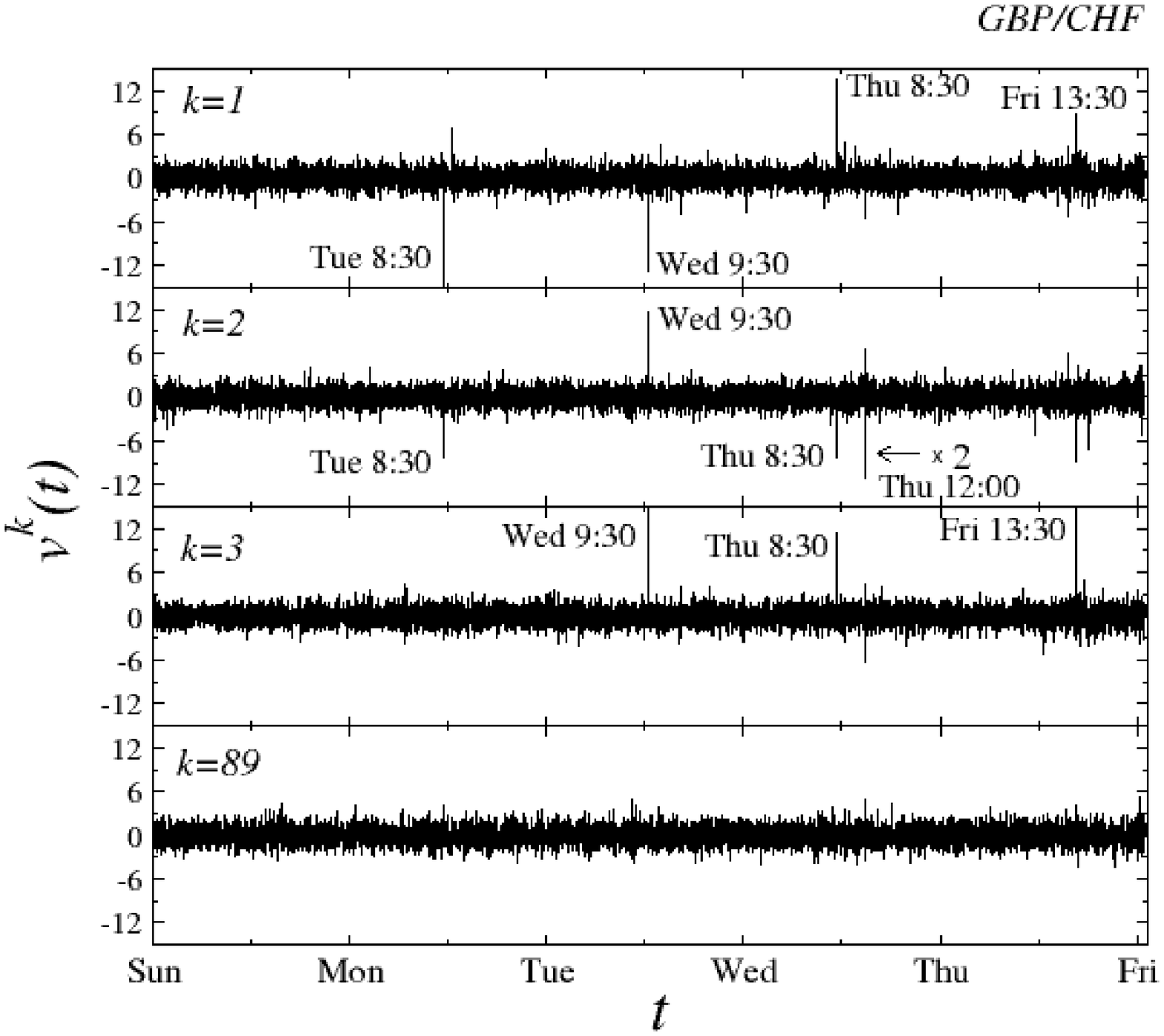}

\vspace{-0.5cm}
\hspace{0.0cm}
\epsfxsize 9cm
\epsffile{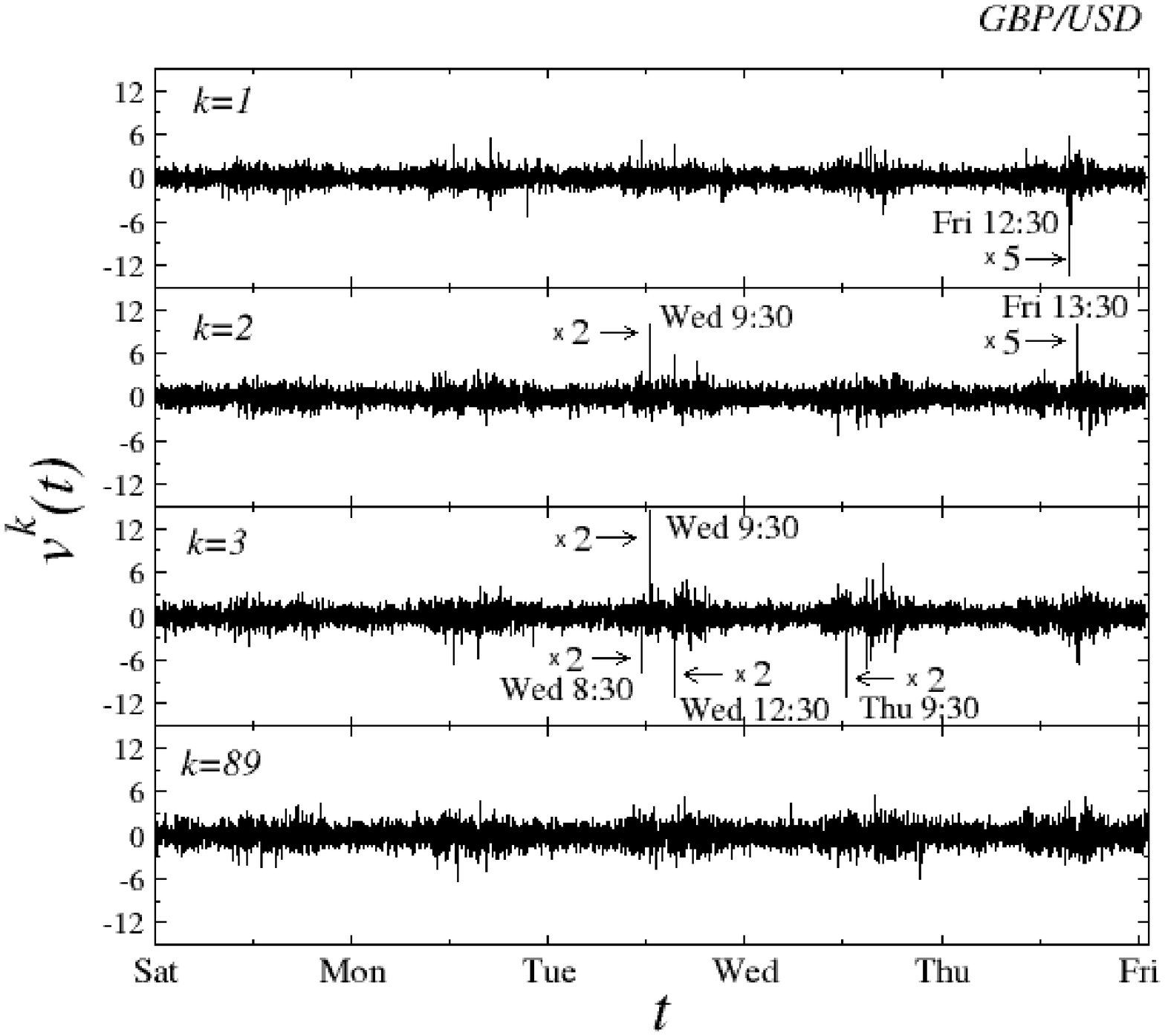}
\hspace{-1.5cm}
\epsfxsize 9cm
\epsffile{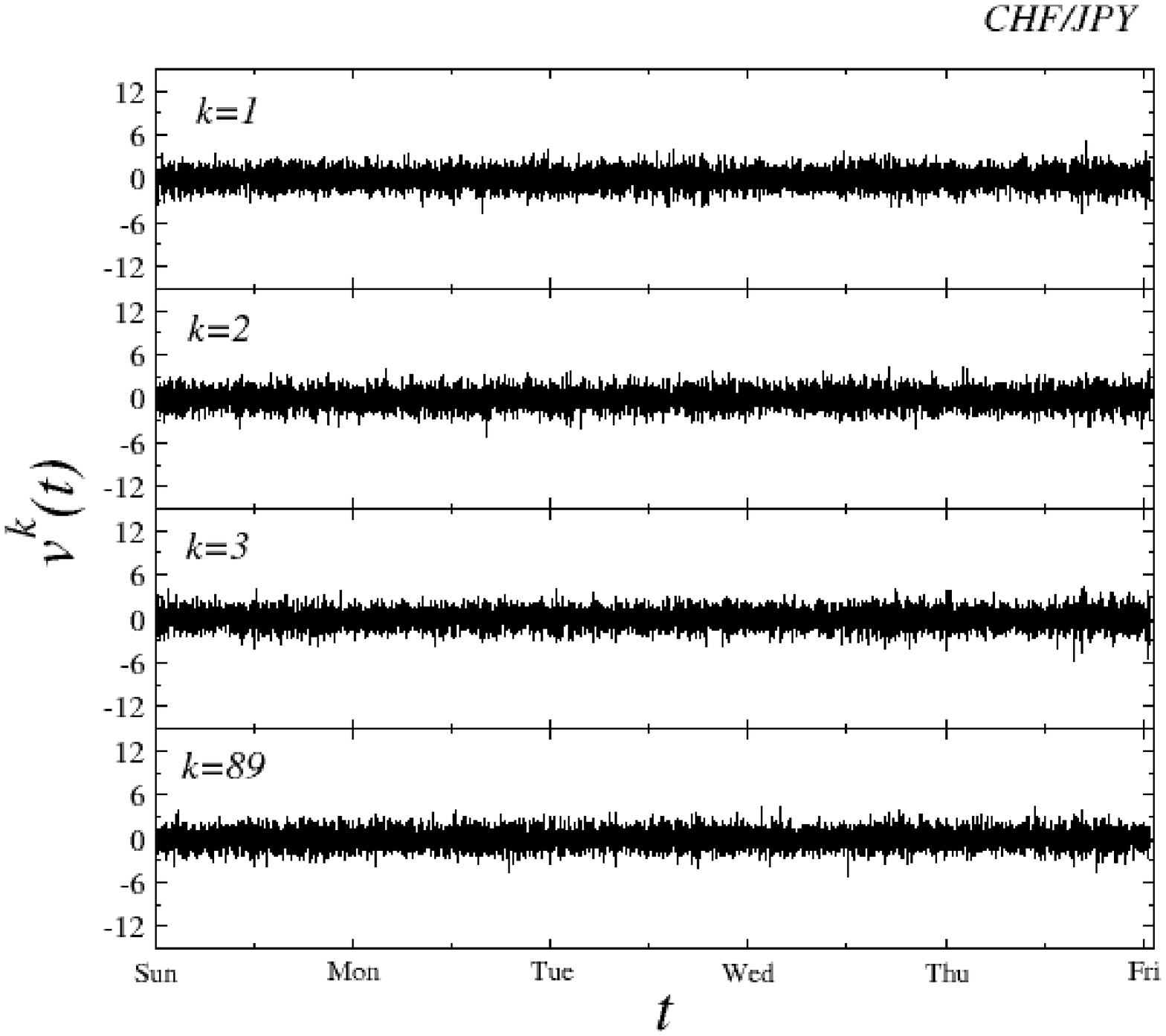}

\vspace{-0.5cm}
\hspace{0.0cm}
\epsfxsize 9cm
\epsffile{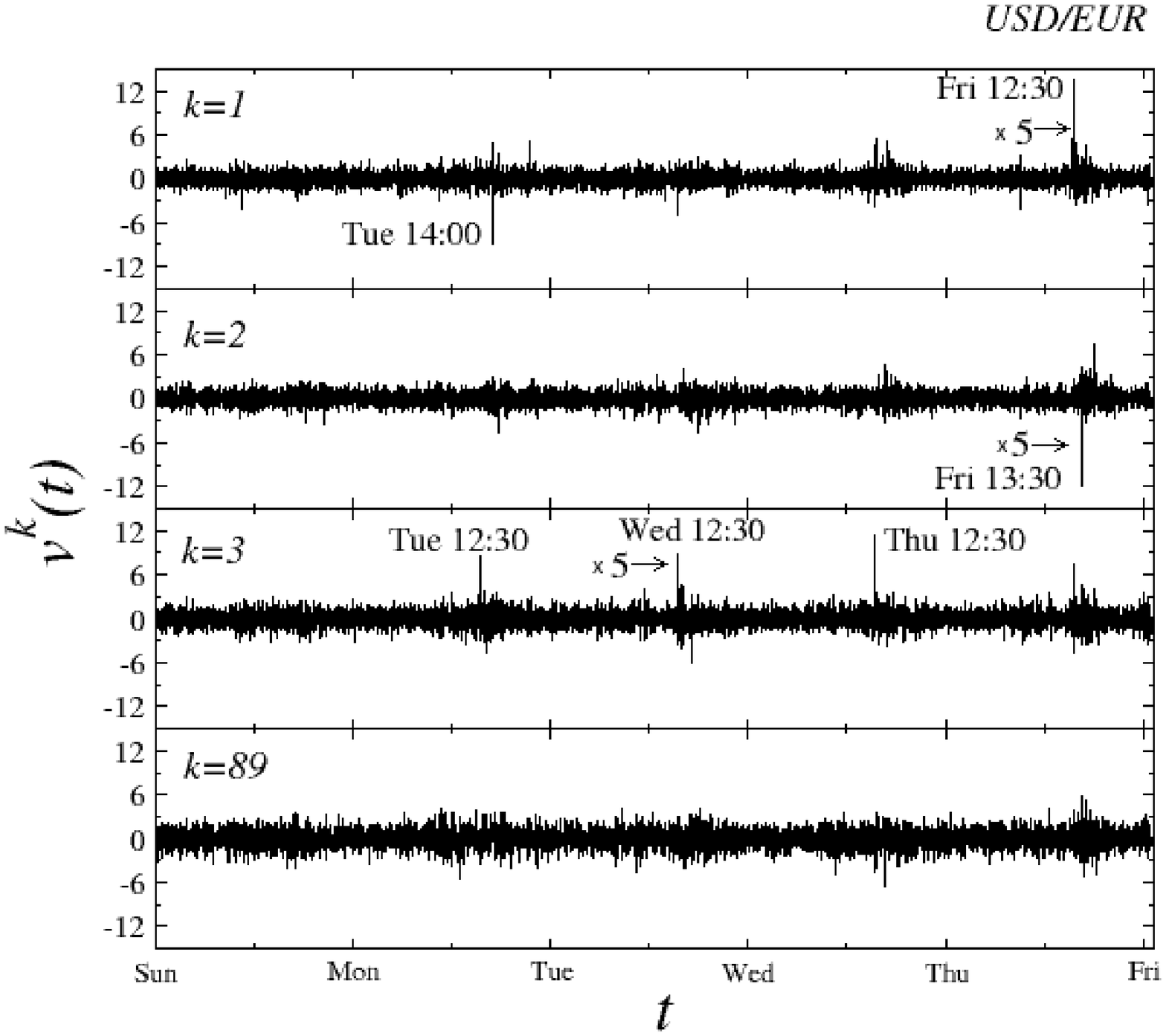}
\hspace{-1.5cm}
\epsfxsize 9cm
\epsffile{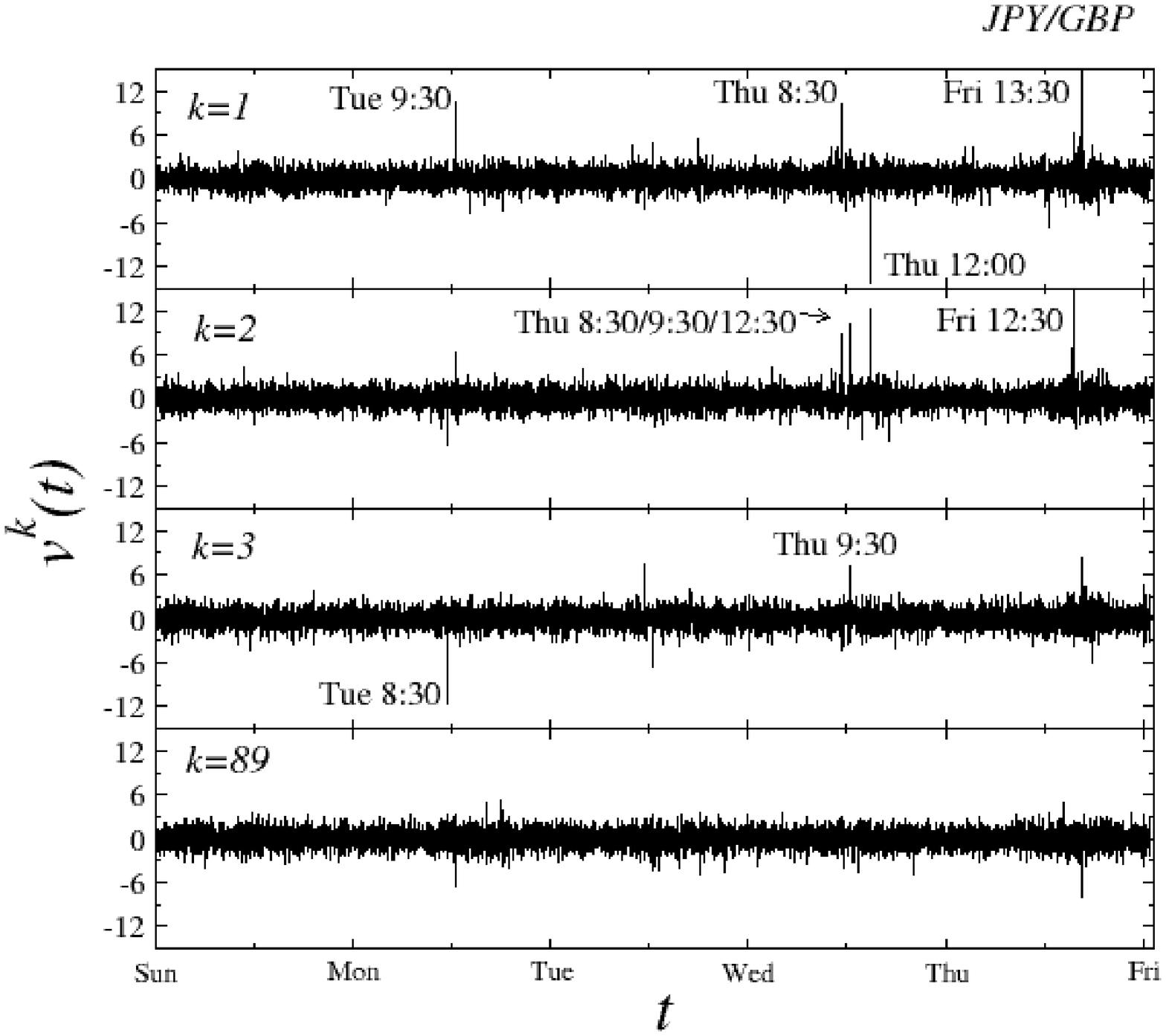}
\caption{Eigensignals corresponding to three largest eigenvalues ($k=1,2,3$) and a typical eigenvalue ($k=89$) of the correlation matrices calculated from the intra-week time series of returns for the GBP-USD-EUR (left) and GBP-CHF-JPY (right) currency triangles. Some largest fluctuations have been suppressed by the indicated factors in order to be fitted into the graphs. Hours are expressed in GMT.}
\label{fig.8}
\end{figure}

More insight into this issue can be gained by looking at the corresponding eigensignals as defined by Eq.~(\ref{eigensignal}). For each exchange rate four such eigensignals are shown in Fig.~8: the ones corresponding to the first three largest eigenvalues and the fourth one which corresponds to an eigenvalue that is embedded deeply $(k=89)$ in the spectrum. Indeed, the cases of the GBP/USD and of the USD/EUR exchange rates look most spectacular. In both cases the first two eigensignals and even the third one display large outstanding fluctuations that by a factor of even about 50 surpass the neighbouring ones. Their presence documents an enhanced market activity systematically at the same instants of time during the consecutive weeks. Interestingly, such instants of time are concentrated more before the weekend than just after it. Such special hours are 13.30 GMT and others. What is natural, the eigensignals from the bulk of the spectrum do not display such a kind of activity. Another fact that deserves special attention is the EUR/GBP exchange rate. Even though belonging to the same triangle, the dynamics is equally smooth for all the eigensignals. One may hypothesize that this indicates a different mechanism that governs dynamics of the exchange rates within this pair as compared to the GBP/USD and USD/EUR. It seems that in a world in which currency trade is dominated by GBP/USD and USD/EUR, the complementary rate EUR/GBP can in the first approximation be considered only as a spectator adjusting its value to the behaviour of the in-play rates. This receives additional support from the fact, that on time scales longer than $\Delta t=1$ min. considered here, the time series of EUR/GBP has fluctuations resembling the ones corresponding to the other exchange rates. From the same perspective, the dynamics within the GBP-CHF-JPY triangle looks more tranquil. Yet, within the JPY/GBP and GBP/CHF pairs one also sees the outlying fluctuations (though relatively smaller than in the previous case) at the well recognizable instants of time during the week period.

\section{Multi-fractal characteristics}

At present, the most compact frame to globally grasp the whole richness of structures and correlations as identified above is - if applicable - in terms of the multifractal spectra~\cite{Halsey1986}. The presence of the long-range nonlinear power law temporal correlations, possibly accompanied by the non-Gaussian character of fluctuations, constitute the necessary - likely not
sufficient however - ingredients in this respect~\cite{Drozdz2009}. Furthermore, by now there exists quite convincing collection of evidence~\cite{Fisher1997,Matia2003,Oswiecimka2005,Kwapien2005} that the financial dynamics often carries signatures of multifractality. In this Section we therefore examine the multifractal characteristics of all the exchange rates considered in the previous Sections.

The Multifractal Detrended Fluctuation Analysis (MFDFA)~\cite{Kantelhardt2002} is the most efficient practical method to quantify multifractality in the financial time series~\cite{Oswiecimka2006}. In MFDFA for an ${x(t_i)}_{i=1,...,T}$ discrete signal one starts with the signal profile $Y(j) = \sum_{i=1}^j{(x(i)-\langle x \rangle)}, \ j = 1,...,T$, where $\langle...\rangle$ denotes averaging over all $i$'s. Then one divides the $Y(j)$ into $M_n$ non-overlapping segments of length $n$ ($n < T$) starting from both the beginning and the end of the signal ($2 M_n$ segments total). For each segment a local trend is estimated by fitting an $l$th order polynomial $P_{\nu}^{(l)}$, which is then subtracted from the signal profile. For the so-detrended signal a local variance $F^2(\nu,n)$ in each segment $\nu$ is calculated for each, from $n_{min}$ to $n_{max}$, scale variable $n$. Finally, by averaging $F^2(\nu,n)$ over all segments $\nu$ one calculates the $r$th order fluctuation function:
\begin{equation}
F_r(n) = \bigg\{ \frac{1}{2 M_n} \sum_{\nu=1}^{2 M_n} [F^2(\nu,n)]^{r/2} \bigg\}^{1/r},
\label{ffunction}
\end{equation}
where $r \in \mathbf{R}$. The relevant power law behavior of the fluctuation function reads:
\begin{equation}
F_r(n) \sim n^{h(r)},
\label{fpower}
\end{equation}
where $h(r)$ is a generalized Hurst exponent. For simple fractals $h(r)= {\rm const}$. If $h(r)$ in addition depends on $r$, the signal is multifractal. Then the singularity spectrum $f(\alpha)$~\cite{Halsey1986} can be calculated as:
\begin{equation}
f (\alpha)=r [\alpha - h (r)] + 1,
\label{falpha}
\end{equation}
where $\alpha=h(r)+r h'(r)$ is the singularity strength. Equivalently, with the commonly used scaling exponent
\begin{equation}
\tau (r) = r h(r) - 1,
\label{tau}
\end{equation}
the singularity strength is expressed as
\begin{equation}
\alpha = {d \tau (r) \over dr}.
\label{alphasing}
\end{equation}

\begin{figure}
\hspace{2.0cm}
\epsfxsize 12cm
\epsffile{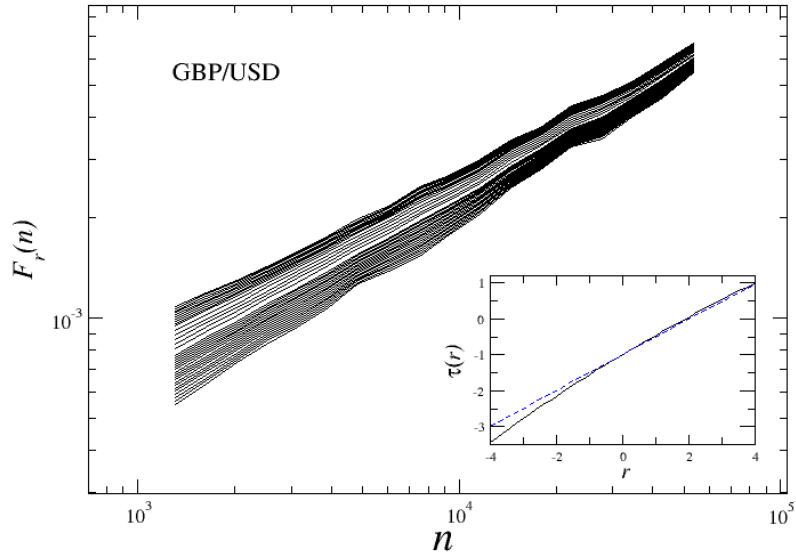}
\caption{(Main) Fluctuation function $F_r(n)$ for an exemplary time series of returns (USD/EUR). Approximate scaling relation for about two decades of $n$ can easily be seen. (Inset) Multifractal spectrum $\tau(r)$ for the same time series (solid line) together with its counterpart for a randomly shuffled time series (dashed line).}
\label{fig.9}
\end{figure}

In ref.~\cite{Drozdz2009} some results were shown concerning requirements to reliably determine the multifractal spectra such that potentially spurious effects are eliminated. This in particular concerns the length of the time series and the quantitative characteristics of the temporal correlations that determine the size of the scaling intervals. Since, as shown above, the FX time series develop heavy tails, the range of the index $r$ needs to be appropriately narrow; we consistently thus choose $r \in [-4,4]$. The detrending polynomial $P_{\nu}^{(l)}$ used is of the second order which, as usual in this kind of analysis, proves to be an optimal choice. An example of $F_r(n)$ for the GBP/USD returns is shown in Fig.~9. For the other exchange
rate returns the overall $F_r(n)$ picture looks qualitatively similar. The scaling of the fluctuation functions $F_r(n)$ is quite convincing and the scaling indices $h(r)$ depend on $r$. This, as is shown in the inset to the same Fig.~9, results in a concave shaped $\tau(r)$ which is characteristic to conventional multifractals. The reference dashed line in this inset represents $\tau(r)$ calculated from the randomized original time series of GBP/USD returns, i.e., by randomly shuffling the data points - a procedure which entirely destroys the temporal correlations. From the corresponding linear dependence of $\tau(r)$ one straightforwardly identifies a monofractal with all the strength concentrated at $\alpha = 0.5$.

\begin{figure}
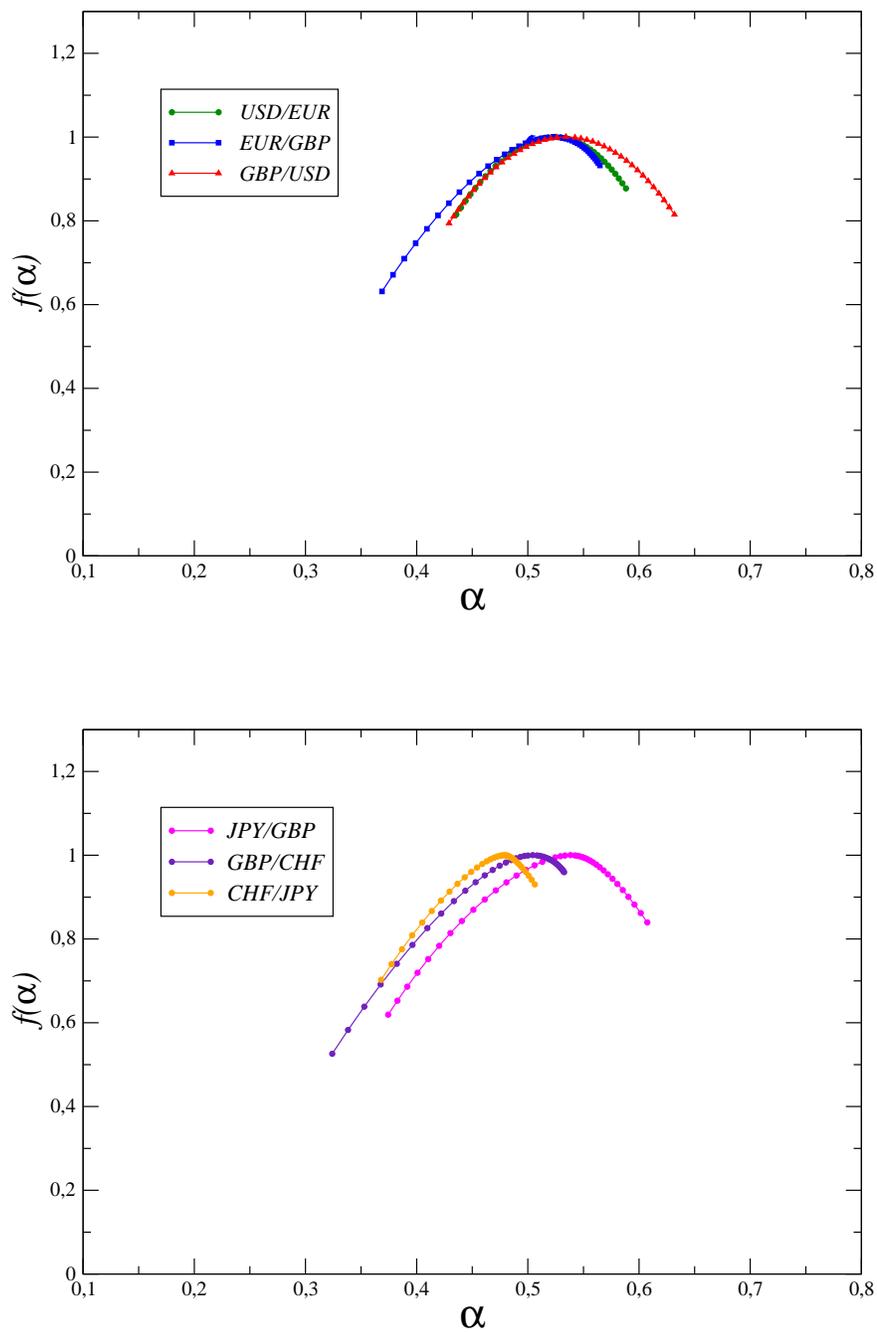

\hspace{2.0cm}
\epsfxsize 11.5cm
\epsffile{fig10a.eps}

\vspace{1.5cm}
\hspace{2.0cm}
\epsfxsize 11.5cm
\epsffile{fig10b.eps}
\caption{Singularity spectra $f(\alpha)$ for time series of returns corresponding to exchange rates between currencies forming the triangles: EUR-GBP-USD (top) and GBP-CHF-JPY (bottom).}
\label{fig.10}
\end{figure}

Singularity spectra $f(\alpha)$ calculated for all the exchange rates are presented in Fig.~10. They are all multifractal with the widths ranging from ca. 0.15 (USD/EUR) to ca. 0.25 (GBP/CHF, CHF/JPY, EUR/GBP) with the maxima located at around $\alpha \approx 0.5$, similarly as for the typical stock market cases. The dispersion of the maxima of $f(\alpha)$ is however larger within the GBP-CHF-JPY triangle than within EUR-GBP-USD. An even more interesting difference is seen in the shape of $f(\alpha)$ for different exchange rates. Majority of them develop an asymmetric $f(\alpha)$ with the distortion somewhat towards the shapes characteristic to bifractals~\cite{Drozdz2009}. The most beautiful and symmetric shape - like the model binomial cascade~\cite{Oswiecimka2006} - is developed by the GBP/USD (London - New York 'cable' connection) and to a lesser extent by the USD/EUR exchange rates. Interestingly, these two are the leading and the most traded exchange rates and the previously seen graphs illustrating their return fluctuations look the most 'erratic' among all the exchange rates considered here. A trace of similar effect can be seen in the second, GBP-CHF-JPY triangle. There, the most symmetric shape of $f(\alpha)$ corresponds to the leading JPY/GBP exchange rate. The degree of symmetry can also be seen to go in parallel with the strength of the volatility autocorrelation (Fig.~5). Stronger volatility autocorrelation corresponds to the more symmetric shape of $f(\alpha)$. These effects to some extent resemble the situation encountered in the human electrocardiogram. There the most healthy and at the same time the most 'erratic' case generates the widest and the most symmetric singularity spectrum~\cite{Ivanov1999}.

\begin{figure}
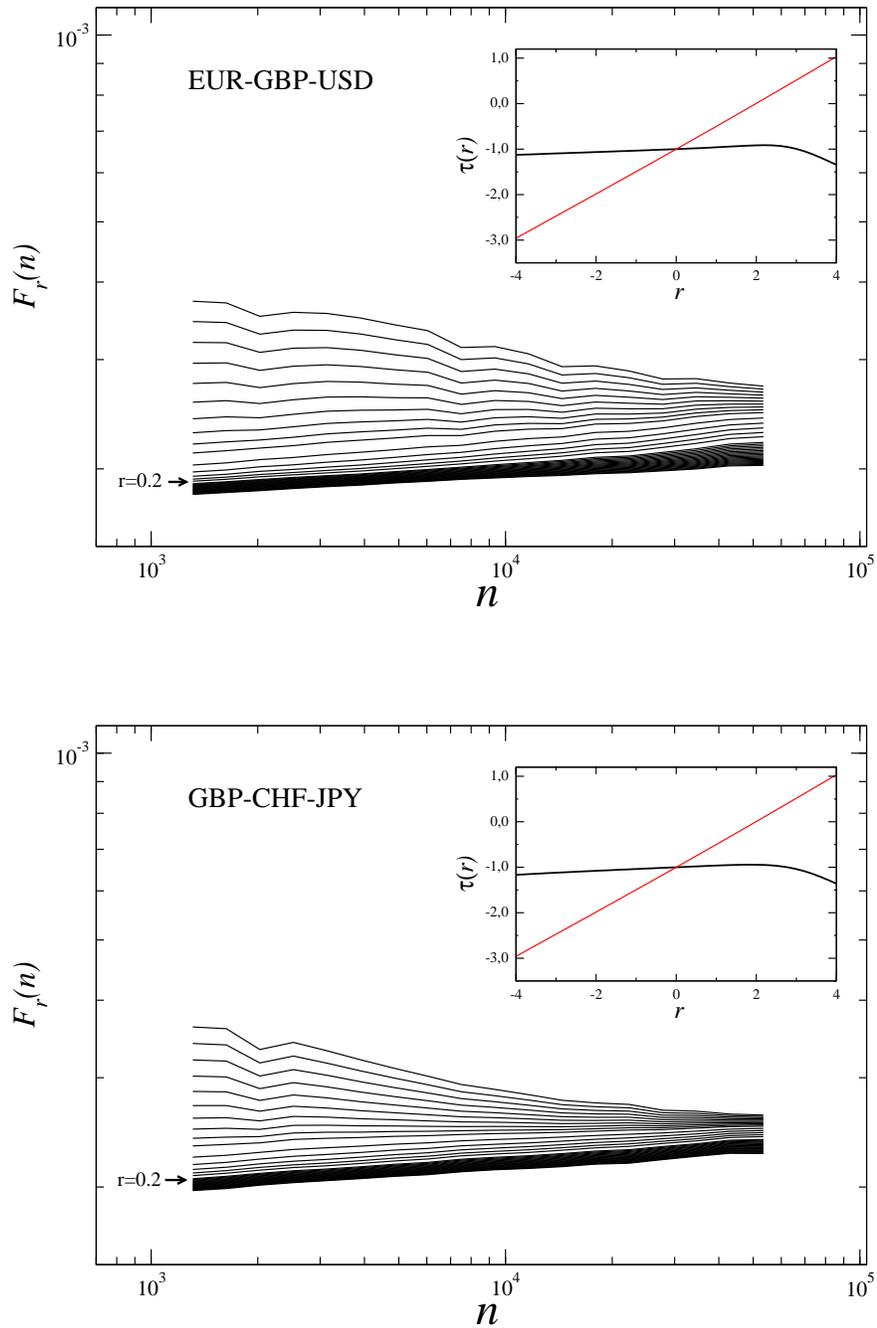

\hspace{2.0cm}
\epsfxsize 11.5cm
\epsffile{fig11a.eps}

\vspace{1.5cm}
\hspace{2.0cm}
\epsfxsize 11.5cm
\epsffile{fig11b.eps}
\caption{Fluctuation function $F_r(n)$ (main) and multifractal spectrum $\tau(r)$ (inset) for time series of residual returns $g^{\bigtriangleup}$ correspodning to the EUR-GBP-USD triangle (top) and the GBP-CHF-JPY triangle (bottom). In both main panels the fluctuation functions are shown for $r \in [-4,4]$ with a step of 0.4. An anomalous scaling behaviour of $F_r(n)$ with negative slope is seen in both main panels for large positive values of $r$. Note also a small dispersion in the slope of $F_r(n)$ seen for the EUR-GBP-USD triangle and negative values of $r$ (top). As a reference, both insets also present $\tau(r)$ for randomly shuffled time series (red line).}
\label{fig.11}
\end{figure}

In contrast to the proper exchange rate returns, we do not observe such conventionally interpretable multifractal characteristics for time series of the residual returns $g^{\bigtriangleup}$. Complexity of the processes underlying such signals can be assessed from the scale $n$-parameter dependence of the fluctuation function $F_r(n)$ for different values of $r$. The result based on calculation within the same range of the parameters $r$ as before is shown in Fig.~11a for the EUR-GBP-USD triangle and in Fig.~11b for the GBP-CHF-JPY triangle. While for the individual values of $r$ the fluctuation functions $F_r(n)$ clearly behave scale free to a similar accuracy as in Fig.~9, the $r$-dependence of the corresponding scaling indices $h(r)$ is significantly different. For the positive values of $r$, with increasing their values, the slope (in the log-log scale) of $F_r(n)$ systematically redirects its orientation such that $h(r)$ becomes negative. For the negative values of $r$, on the other hand, the slope of $F_r(n)$ - and thus $h(r)$ - almost does not depend on $r$ which signals a monofractal character of the small fluctuations as these are predominantly filtered by the negative parameters $r$. The resulting $\tau(r)$ is shown in the inset and can be seen to have a profoundly different functional form as compared to the original returns case of Figs.~9 and 10. Still, by randomly shuffling the residual returns time series one obtains the same monofractal form of $\tau(r)$ (dashed lines in the insets to Fig.~11a and 11b) as before when the series of returns were shuffled. This signals that the presently observed anomalous functional form of $\tau(r)$ for the residual returns $g^{\bigtriangleup}$ is primarily encoded in the specific form of the temporal correlations in $g^{\bigtriangleup}$.

\begin{figure}
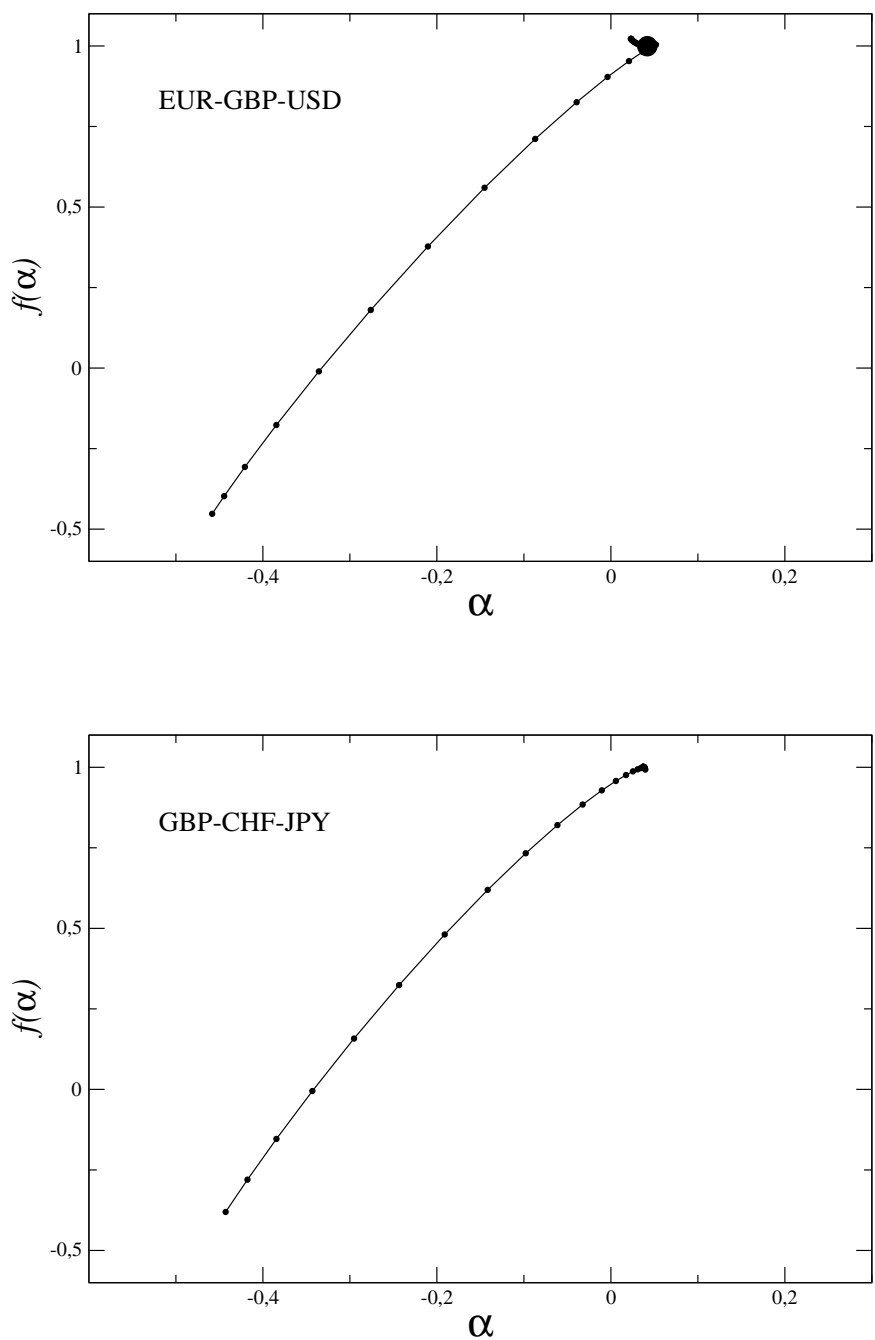

\hspace{2.0cm}
\epsfxsize 11.5cm
\epsffile{fig12a.eps}

\vspace{1.5cm}
\hspace{2.0cm}
\epsfxsize 11.5cm
\epsffile{fig12b.eps}
\caption{Singularity spectra $f(\alpha)$ for time series of residual returns $g^{\bigtriangleup}$ corresponding to the triangles: EUR-GBP-USD (top) and GBP-CHF-JPY (bottom). The aggregation of data points near $\alpha \approx 0.04$ observed for EUR-GBP-USD is related to a small dispersion in the slope of $F_r(n)$ for negative values of $r$ in the top panel of Fig.~11.}
\label{fig.12}
\end{figure}

The singularity spectra $f(\alpha)$ that correspond to the above two cases are shown in Fig.~12. They develop essentially only the left wing which corresponds to the positive values of $r$. Somewhat related ``left-sided'' multifractals have in fact already been considered in the literature~\cite{mandelbrot90a,mandelbrot90b,lee88} in applications to diffusion limited aggregates (DLA) and to fully developed turbulence. This may signal further analogies between the FX dynamics and the physics of turbulence in accord and giving more arguments in favor of the conjecture put forward in ref.~\cite{ghashghaie96}. Furthermore, $f(\alpha)$ extends to the negative values of the singularity exponents $\alpha$ where, at the edge, $f(\alpha)$ even assumes the negative values. To our knowledge such an anomalous form of multifractality has never been identified before in the context of the financial dynamics. However even such a possibility appears to be implicitly involved in the Mandelbrot considerations~\cite{mandelbrot74} on the fluid dynamics and already explicitly in more recent statements on the issue of negative critical dimensions~\cite{mandelbrot03}. This latter extended study has been motivated by a rigorous demonstration~\cite{duplantier99} of the presence of negative fractal quantities for the (conformal invariant) harmonic measure around a number of incipient percolation clusters. A related indication is that the ``multifractal anomalies'' arise when the system under study behaves canonically - in the statistical physics sense - instead of microcanonically. Illustrated by means of the binomial cascade this extends to the situation such that the sum of partitions at each recursion is not preserved exactly but only in the average. Several quantitative characteristics seen above indicate that the dynamics associated with the constraint imposed by the FX triangle rule belongs to this category of phenomena. In the real FX dynamics the triangle relation expressed by the Eq.~\ref{triangle} is obeyed also only in the average. A complementary interpretation of the negative fractal dimensions is that they describe the missing fluctuations - therefore typically large and thus filtered by the positive parameters $r$ - in a studied finite size sample. Such fluctuations are thus expected to come into view in another realization of a finite size sample from the same ensemble.

\section{Cross-correlations}

It is well-known that time series of returns of different assets traded on the same market are typically cross-correlated. This holds true also for the currency market~\cite{dacorogna01,drozdz07}. Unlike other studies before, here we analyze a specific case of correlations between the exchange rates coupled by the triangle rule. We expect that deviations from the perfect triangle relation can be observed not only by means of the residual returns distributions (Fig.~3) but also by means of the eigenvalue spectra of the correlation matrix calculated from the triples of time series corresponding to currency triangles. We can exploit here the fact that our time series were recorded simultaneously. We follow the same procedure of constructing a correlation matrix as above, but now we consider the complete time series of length $T=1,703,520$. For a few different choices of the time scale $\Delta t$, we create two matrices of size $3 \times 3$, each for one currency triangle. Due to Eq.~(\ref{triangle}), the triangle rule, if fulfilled, implies that ${\bf C}$ has only two non-zero eigenvalues whose sum satisfies the condition: $\lambda_1+\lambda_2 = 3$. Existence of $\lambda_3 > 0$ would thus mean a possibility of a triangular arbitrage.

\begin{figure}
\hspace{-0.5cm}
\epsfxsize 9.0cm
\epsffile{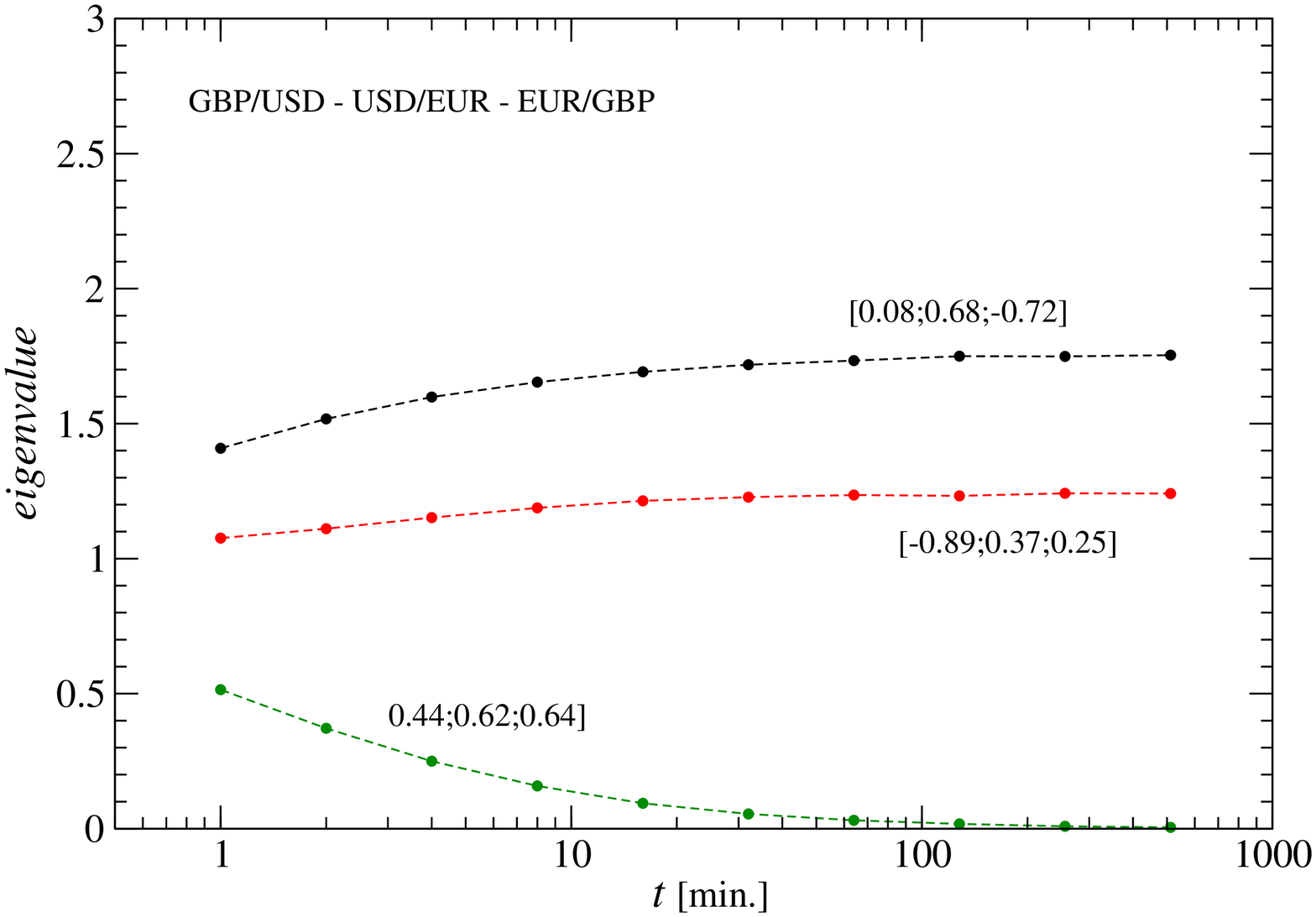}
\hspace{-1.0cm}
\epsfxsize 9.0cm
\epsffile{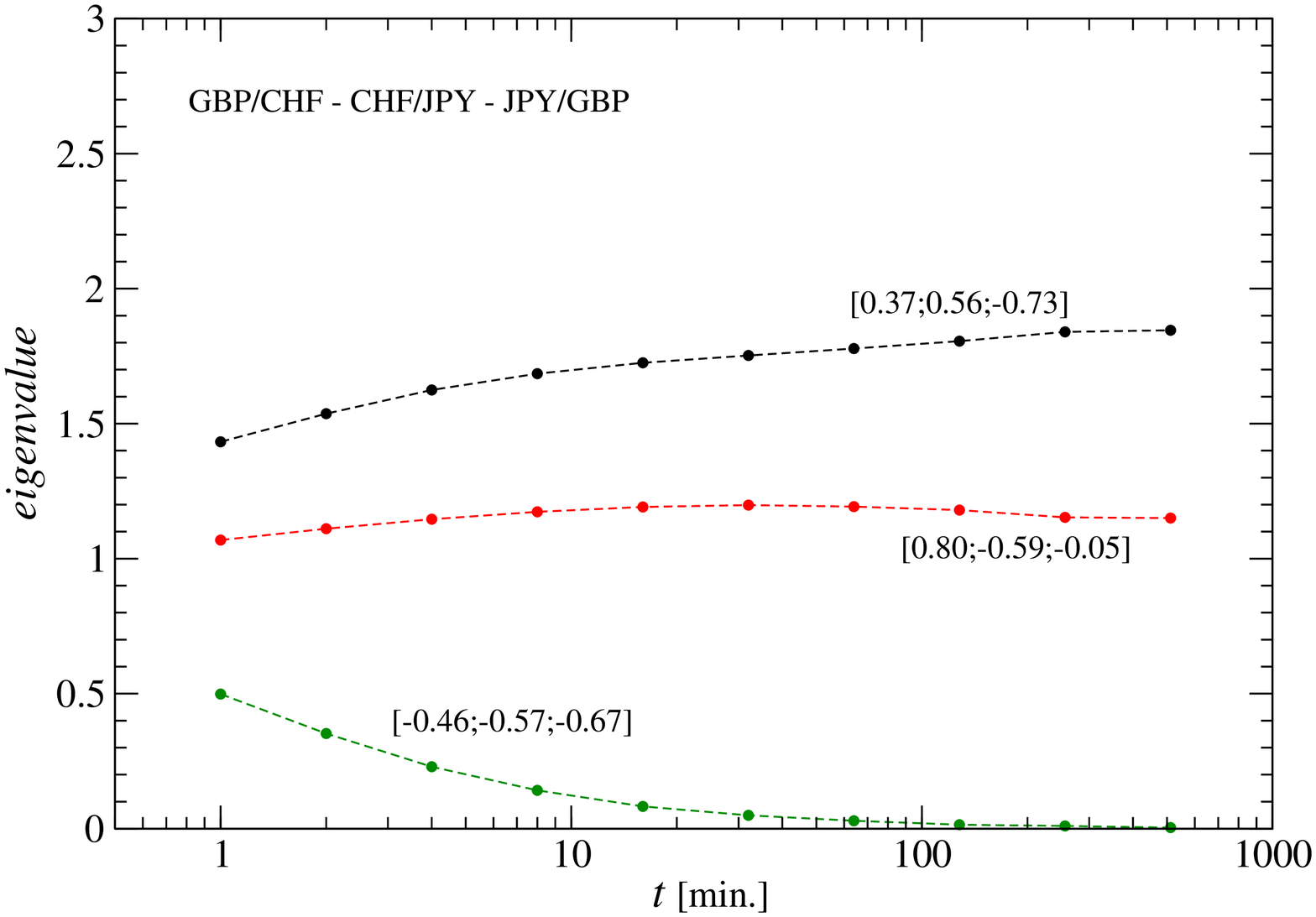}

\vspace{0.0cm}
\hspace{-0.5cm}
\epsfxsize 9.0cm
\epsffile{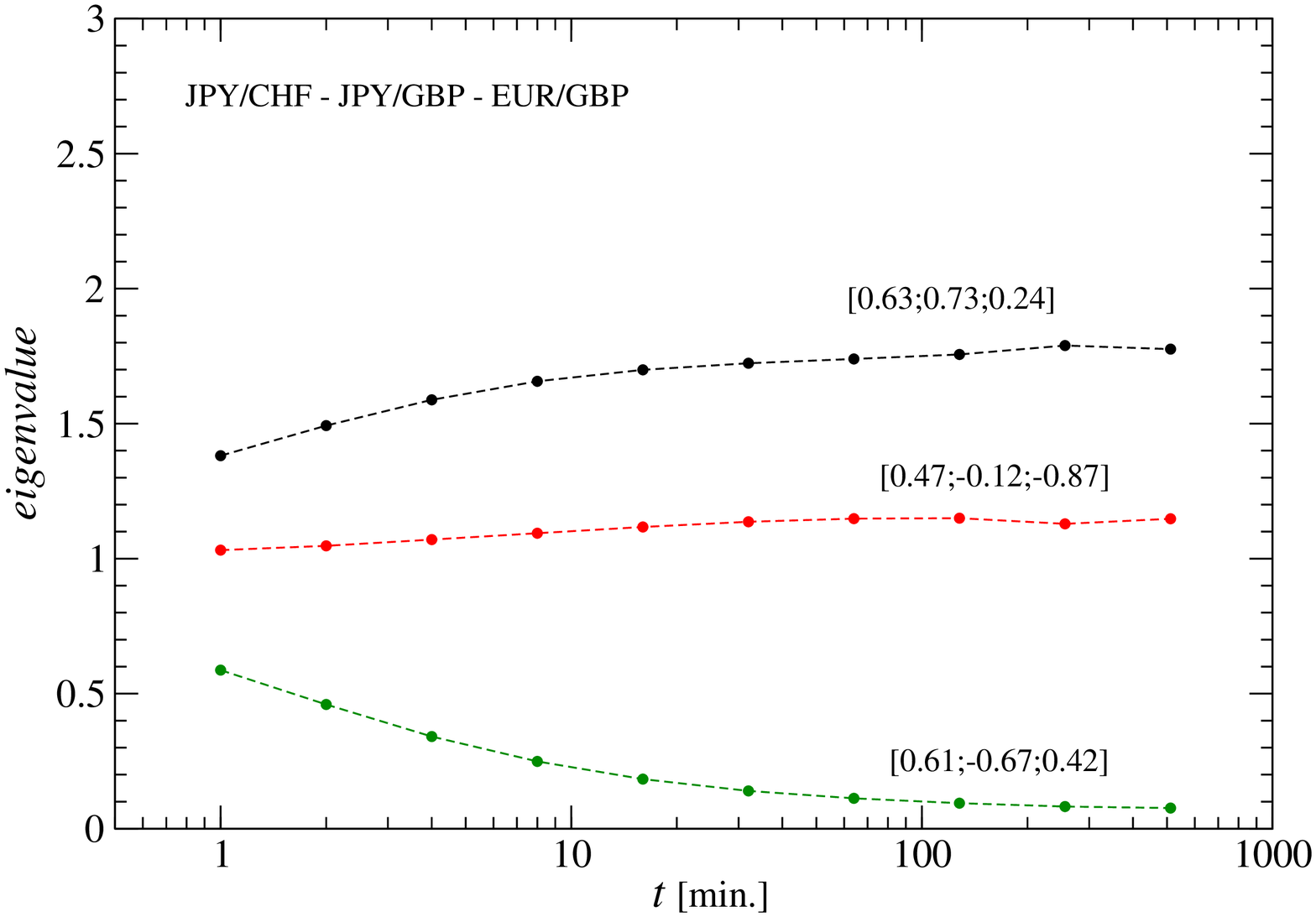}
\hspace{-1.0cm}
\epsfxsize 9.0cm
\epsffile{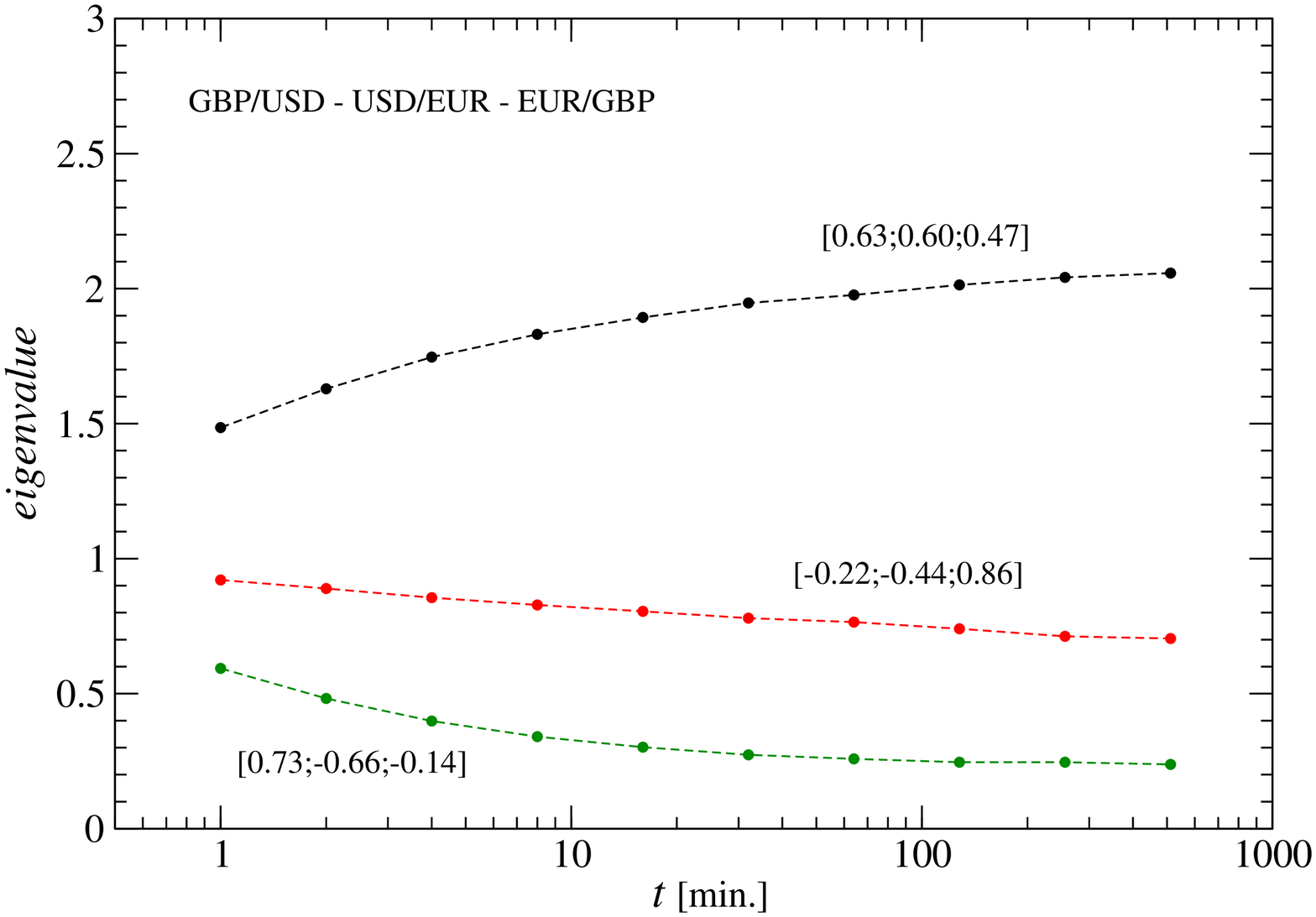}
\caption{(Top) Eigenvalues of the correlation matrix calculated for three time series of returns corresponding to three exchange rates forming two triangles: EUR-GBP-USD (left) and GBP-CHF-JPY (right). Each eigenvalue is shown as a function of the time scale $\Delta t$. The corresponding eigenvector components for the longest time scale considered here ($\Delta t=512$ min.) are printed in square brackets. (Bottom) The same for three exchange rates which do not form a triangle: JPY/GBP, JPY/CHF, EUR/GBP (left), and GBP/USD, USD/EUR, EUR/GBP (right).}
\label{fig.13}
\end{figure}

In Fig.~13 top panels show functional dependence of the eigenvalues of ${\bf C}$ on $\Delta t$ for the two considered currency triangles. For the shortest $\Delta t=1$ min., the data clearly does not comply with Eq.~(\ref{triangle}) and ${\bf C}$ has three non-zero eigenvalues in both cases. Although possibilities of the triangular arbitrage, with the today's computer trading, are not expected to last longer than a fraction of a second, in the correlation matrix representation their trace can be seen clearly on much longer time scales. This is because all the exchange rates were sampled precisely at the same time and, thus, the inconsistencies in exchange rates could not be consumed yet. Since these inconsistencies, as regards their absolute magnitude, are the same no matter which time scale one considers, their relative magnitude should gradually decline with increasing $\Delta t$ (and, therefore, with increasing variance of the unnormalized returns). This effect should manifest itself by a declining value of $\lambda_3(\Delta t)$. Due to the fact that trace of ${\bf C}$ is independent of $\Delta t$, decreasing the level of $\lambda_3$ must be associated with an increase of the two remaining eigenvalues. This resembles the well-known Epps effect observed on the stock markets~\cite{epps79}. In fact, even the time scales at which $\lambda_1(\Delta t)$ saturates (50-100 min.) are roughly the same as those found for stocks.

The same effect can in general be observed for any triples of the exchange rates not neccessarily forming a triangle, as it is documented by two examples shown in bottom panels of Fig.~13, where the exchange rates are formed from 4 currencies and, thus, do not constitute any cycle. The only quantitative difference between the eigenvalues of the triangles and the ``non-triangles'' is in the asymptotic magnitude of $\lambda_3(\Delta t)$, which in the former case is zero and in the latter case is small but positive. One therefore sees that the triangle rule implies that the fully-developed couplings among the involved exchange rates are associated with a zero mode of ${\bf C}$.

The origin of the Epps effect in the forex market is likely to be similar to its counterpart for the stock market: a finite speed of information spreading among the assets. One possible source is a lack of transaction synchronicity on different assets, which introduces noise-like effects on their correlated evolution~\cite{kwapien04,toth09}. However, this non-synchronicity of trading is probably not the unique cause of the Epps effect in the currency market: the trading frequency on this market is much higher than its counterpart on the stock markets, yet the time scales of saturation are comparable on both market types. This suggests that some other factor may play an important role in the development of correlations. Indeed, other sources of the Epps effect have already been proposed and may be relevant here like the microstructure noise and the discretization error~\cite{zhang10}. However, their influence in this case has yet to be assessed.

\section{Summary}

In this paper we analyzed time series of currency exchange rate returns for the two triples of currencies forming triangles: EUR-GBP-USD and GBP-CHF-JPY. Market efficiency requires that cycling through currencies in such triangles must not be profitable except very short time scales, which is reflected in the triangle rule. For the original FX time series we find that the main statistical properties of the corresponding returns - their distributions, temporal correlations, and multifractality - are qualitatively similar to those found for other markets. However, we found also some quantitative differences between properties of different exchange rates which may reflect their different significance in the world currency exchange system. We also studied the residual signals consisting of short-time deflections from the perfect no-arbitrage condition. A related interesting observation is that while the proper exchange rate returns are well modeled by the $q$-Gaussian distributions, the residual returns develop disproportionately heavier tails.

Among the most illuminating views is the one that can be obtained after diagonalizing the correlation matrices constructed from time series representing different weeks and calculating the corresponding eigensignals, i.e. independent components of dynamics associated with repeatable patterns of activity. Eigensignals carry system-specific information if one can identify large fluctuations which can be related to some periodic external perturbation of the market (e.g. economical news releases). It occurs that such fluctuations are clearly visible already on 1 min. time scale for heavily-traded cross-rates like USD/EUR and GBP/USD, but only on longer time scales for less frequently traded rates like EUR/GBP. The same refers to the second triangle, in which more popular rates: JPY/GBP and GBP/CHF have more characteristic eigensignals at 1 min. time scale, while the less popular CHF/JPY rate has more universal (noisy) eigensignals. We argued that this effect reflects the fact that less popular exchange rates play a passive role, tuning their values according to changes in dominant rates as demanded by the triangle rule.

A parallel effect is related to the different shapes of the singularity spectra for different currency pairs. In this respect the most symmetric $f(\alpha)$ spectrum is observed for the GBP/USD pair, while other currency pairs have spectra which are more asymmetric, especially those from the GBP-CHF-JPY triangle. Even more intriguing are the signatures of negative singularity exponents and negative singularity spectra for the triangle residual returns. This opens an exciting direction for further investigations towards perhaps establishing closer analogy between the FX dynamics and the phenomenon of turbulence.

We also found that some inefficiency of the market allowed for extremely short time scales leads to the emergence of the Epps effect, i.e. an increase of couplings between different exchange rates from the same triangle, if going from shorter to longer time scales. Our result indicates that, from a point of view of returns, the influence of market inefficiency on cross-correlations among exchange rates can be neglected on time scales longer than, roughly, an hour. An increase of coupling with similar characteristics time scales involved is in fact observed for any triples of exchange rates not necessarily forming a triangle.

\section{Appendix: $q$-Gaussian distribution}

\setcounter{equation}{3}

The $q$-Gaussian distribution is defined by~\cite{Tsallis1995}:
\begin{equation}
p\left( x\right) =\mathcal{N}_{q}\,e_{q}^{-\mathcal{B}_{q}\left( x-\bar{\mu}_{q}\right) ^{2}},
\end{equation}
\setcounter{equation}{15}
where
\begin{eqnarray}
\mathcal{N}_{q}=\left\{
\begin{array}{ccc}
\frac{\Gamma \left[ \frac{5-3q}{2-2q}\right] }{\Gamma \left[ \frac{2-q}{1-q} \right] }\sqrt{\frac{1-q}{\pi }\mathcal{B}_{q}} & ~~{\rm for}  & q<1 \\[3mm] \frac{\Gamma \left(\frac{1}{q-1}\right)}{\Gamma \left(\frac{3-q}{2 (q-1)}\right) \sqrt{\frac{\pi}{(q-1) \mathcal{B}_q}}} & ~~{\rm for} & 1<q<3
\end{array}
\right. \\
\bar{\mu} _{q}= \,\int x\frac{\ \left[ p\left( x\right) \right] ^{q}}{\int \left[ p\left( x\right) \right] ^{q}dx}\ dx\equiv \left\langle x\right\rangle _{q}, \ \ \ \ \mathcal{B}_{q}=\left[ \left( 3-1\right) \,\bar{\sigma}_{q}^{2}\right] ^{-1}
\nonumber
\end{eqnarray}
and $e_q^x$ denotes the $q$-exponential function
\begin{equation}
e_{q}^{x}=\left[ 1+\left( 1-q\right) \,x\right] ^{ \frac{1}{1-q}}.
\label{eq}
\end{equation}

For $q>1$ this distribution asymptotically $(x \gg 1)$ develops a power law form $p(x)\sim x^{\frac{2}{1-q}}$. In particular, for $q=3/2$, on the level of the cumulative distribution, it recovers the inverse cubic power law. This is a particularly useful aspect of the functional form expressed by Eq.~(\ref{px}) because it at the same time provides a compact form for the probability distribution for any value of $x$.

Instead of directly using Eq.~(\ref{px}) it is more practical to convert it to the cumulative form by defining
\begin{equation}
P_{\pm}(x) =\mp\int_{\pm \infty}^x p(x')dx' \label{Pc},
\end{equation}
where the + and - signs correspond to the right and left wings of the distribution, respectively. By using Eq.~(\ref{px}) one obtains
\begin{equation}
P_{\pm}(x) =\mathcal{N}_q\left(\frac{\sqrt{\pi } ~\Gamma \left(\frac{1}{2} (3-q) ~\beta \right)}{2 ~\Gamma (\beta )~\sqrt{\frac{\mathcal{B}_q}{\beta }}}\pm(x-\bar{\mu}_{q}) \, _2F_1(\alpha ,\beta ;\gamma ;\delta )\right)~,
\label{Pcx}
\end{equation}
where $\alpha=\frac{1}{2}$, $\beta =\frac{1}{q-1}$, $\gamma =\frac{3}{2}$, $\delta =-\mathcal{B}_q(q-1)(\bar{\mu}_{q}-x)^2$ and $_2F_1(\alpha ,\beta ;\gamma ;\delta)$ is the Gauss hypergeometric function:
\begin{equation}
_2F_1(\alpha ,\beta ;\gamma ;\delta) = \sum_{k=0}^\infty {\frac{\delta ^k ~(\alpha)_k~(\beta)_k}{k!~(\gamma) _k}}
\label{2F1}
\end{equation}


\section*{References}

\begin{thebibliography}{99}

\bibitem{data} http://www.forexrate.co.uk

\bibitem{Aiba2003} Aiba Y, Hatano N, Takayasu H, Marumo K and Shimizu T 2003 {\it Physica A} {\bf 324} 253

\bibitem{Gorski2008} G\'orski A Z, Dro\.zd\.z S and Kwapie\'n J 2008 {\it Eur.~Phys.~J.~B} {\bf 66} 91

\bibitem{Kwapien2009a} Kwapie\'n J, Gworek S and Dro\.zd\.z S 2009 {\it Acta Phys. Pol.~B} {\bf 40} 175

\bibitem{Kwapien2009b} Kwapie\'n J, Gworek S, Dro\.zd\.z S and G\'orski A Z 2009 {\it J.~Econ. Interact.~Coord.} {\bf 4} 55

\bibitem{Gabaix2003} Gabaix X, Gopikrishnan P, Plerou V and Stanley H E 2003 {\it Nature} {\bf 423} 267

\bibitem{Lux1996} Lux T 1996 {\it Appl.~Financial~Economics} {\bf 6} 463

\bibitem{Gopi1999} Gopikrishnan P, Plerou V, Amaral L A N, Meyer M and Stanley H E 1999 {\it Phys.~Rev.~E} {\bf 60} 5305

\bibitem{Plerou1999} Plerou V, Gopikrishnan P, Amaral L A N, Meyer M and Stanley H E 1999 {\it Phys. Rev. E} {\bf 60} 6519

\bibitem{Drozdz2003} Dro\.zd\.z S, Kwapie\'n J, Gr\"ummer F, Ruf F and Speth J 2003 {\it Acta Phys. Pol. B} {\bf 34} 4293

\bibitem{Rak2007} Rak R, Dro\.zd\.z S and Kwapie\'n J 2007 {\it Physica A} {\bf 374} 315

\bibitem{Matia2002} Matia K, Amaral L A N, Goodwin S and Stanley H E 2002 Phys. Rev. E {\bf 66} 045103

\bibitem{Muller1995} M\"uller U A, Dacorogna M M, Olsen R B, Pictet O V, Schwarz M and Morgenegg C 1995 {\it J.~Banking Finance} {\bf 14} 1189

\bibitem{Drozdz2009} Dro\.zd\.z S, Kwapie\'n J, O\'swi\c ecimka P and Rak R 2009 {\it EPL} {\bf 88} 60003

\bibitem{Tsallis1988} Tsallis C 1988 {\it J.~Stat.~Phys.} {\bf 52} 479

\bibitem{Tsallis1995} Tsallis C, Levy S V F, Souza A M C and Maynard R 1995 {\it Phys.~Rev.~Lett.} {\bf 75} 3589

\bibitem{gillaume97} Gillaume D, Dacorogna M M, Dave R, M\"uller U, Olsen R B and Pictet O V 1997 {\it Fin. Stochastics} {\bf 1} 95

\bibitem{Drozdz2007} Dro\.zd\.z S, Forczek M, Kwapie\'n J, O\'swi\c ecimka P and Rak R 2007 {\it Physica A} {\bf 383} 59

\bibitem{Drozdz1995} Dro\.zd\.z S, Nishizaki S, Speth J and Wambach J 1995 {\it Phys.~Rev.~Lett.} {\bf 74} 1075

\bibitem{goodhart91} Goodhart C A E and Figuoli L 1991 {\it J.~Intern. Money Finance} {\bf 10} 23

\bibitem{bollerslev93} Bollerslev T and Domowitz I 1993 {\it J.~Finance} {\bf 48} 1421

\bibitem{bollerslev94} Bollerslev T and Melvin M 1994 {\it J.~Intern. Econ.} {\bf 36} 355

\bibitem{Ding1993} Ding Z, Granger C W J and Engle R 1993 {\it J.~Empir.~Finance} {\bf 1} 83

\bibitem{Bouchaud2000} Bouchaud J-P 2000 {\it Physica A} {\bf 285} 18

\bibitem{Kwapien2000} Kwapie\'n J, Dro\.zd\.z S and Ioannides A A 2000 {\it Phys.~Rev.~E} {\bf 62} 5557

\bibitem{Kwapien2002} Kwapie\'n J, Dro\.zd\.z S, Gr\"ummer F, Ruf F and Septh J 2002 {\it Physica A} {\bf 309} 171

\bibitem{Marcenko1967} Mar\u cenko V A and Pastur L A 1967 {\it Math.~USSR~Sb.} {\bf 1} 457

\bibitem{Sengupta1999} Sengupta A M and Mitra P P 1999 {\it Phys.~Rev.~E} {\bf 60} 003389

\bibitem{Halsey1986} Halsey T C, Jensen M H, Kadanoff L P, Procaccia I and Shraiman B I 1986 {\it Phys.~Rev.~A} {\bf 33}, 1141

\bibitem{Fisher1997} Fisher A, Calvet L and Mandelbrot B 1997 ``Multifractality of Deutschemark/US dollar exchange rates'', Cowles Foundation Discussion Paper No. 1166

\bibitem{Matia2003} Matia K, Ashkenazy Y and Stanley H E 2003 {\it Europhys.~Lett.} {\bf 61} 422

\bibitem{Oswiecimka2005} O\'swi\c ecimka P, Kwapie\'n J and Dro\.zd\.z S 2005 {\it Physica A} {\bf 347} 626

\bibitem{Kwapien2005} Kwapie\'n J, O\'swi\c ecimka P and Dro\.zd\.z S 2005 {\it Physica A} {\bf 350} 466

\bibitem{Kantelhardt2002} Kantelhardt J W, Zschiegner S A, Koscielny-Bunde E, Bunde A, Havlin S and Stanley H E 2002 {\it Physica A} {\bf 316} 87

\bibitem{Oswiecimka2006} O\'swi\c ecimka P, Kwapie\'n J and Dro\.zd\.z S 2006 {\it Phys.~Rev.~E} {\bf 74} 016103

\bibitem{Ivanov1999} Ivanov P C, Amaral L A N, Goldberger A L, Havlin S, Rosenblum M G, Struzik Z R and Stanley H E 1999 {\it Nature} {\bf 399} 461

\bibitem{mandelbrot90a} Mandelbrot B B 1990 {\it Physica A} {\bf 168} 95

\bibitem{mandelbrot90b} Mandelbrot B B, Evertsz C J G, Hayakava Y 1990 {\it Phys. Rev. A} {\bf 42} 4528

\bibitem{lee88} Lee J, Stanley H E 1988 {\it Phys. Rev. Lett.} {\bf 61} 2945

\bibitem{ghashghaie96} Ghashghaie S, Breymann W, Peinke J, Talkner P and Dodge Y 1996 {\it Nature} {\bf 381} 767

\bibitem{mandelbrot74} Mandelbrot B B 1974 {\it J.~Fluid~Mech.} {\bf 62} 331

\bibitem{mandelbrot03} Mandelbrot B B 2003 {\it J.~Stat.~Phys.} {\bf 110} 739

\bibitem{duplantier99} Duplantier B 1999 {\it Phys. Rev. Lett.} {\bf 82} 3940

\bibitem{dacorogna01} Dacorogna M M, Gen\c cay R, M\"uller U, Olsen R B and Pictet O V, ``An Introduction to High-Frequency Finance'' Academic Press (San Diego, 2001)

\bibitem{drozdz07} Dro\.zd\.z S, G\'orski A Z and Kwapie\'n J 2007 {\it Eur.~Phys.~J.~B} {\bf 58} 499

\bibitem{epps79} Epps T W 1979 {\it J.~Am.~Stat.~Assoc.} {\bf 74} 291

\bibitem{kwapien04} Kwapie\'n J, Dro\.zd\.z S and Speth J 2004 {\it Physica A} {\bf 337} 231

\bibitem{toth09} T\'oth B and Kert\'esz J 2009 {\it Quant.~Finance} {\bf 9} 793

\bibitem{zhang10} Zhang L 2010 ``Estimating covariation: Epps effect, microstructure noise'' {\it J.~Econometrics} in press

\end{thebibliography}
\end{document}